\documentclass[prd,twocolumn,superscriptaddress,floatfix,amsmath,amssymb,amsfonts,longbibliography,nofootinbib]{revtex4-1}

\usepackage{float}

\usepackage{comment}
\usepackage[normalem]{ulem}
\usepackage[english]{babel}
\usepackage{graphicx}
\usepackage{dcolumn}
\usepackage{bm}
\usepackage{blindtext}
\usepackage{verbatim}
\usepackage{mathrsfs}
\usepackage{musicography}
\usepackage{amsmath}
\usepackage{cancel}
\usepackage{physics}
\usepackage{epstopdf}
\usepackage{mathtools}
\usepackage{tensor}
\usepackage{color}
\usepackage[usenames,dvipsnames]{pstricks}
\usepackage{epsfig}
\usepackage{pst-grad} 
\usepackage{pst-plot} 
\usepackage{hyperref}
\usepackage{verbatim}
\usepackage{slashed}

\newcommand{\tb}[1]{\textcolor{black}{#1}}

\newcommand{\mf}{\mathsf}

\newcommand{\ii}{\mathrm{i}}

\newcommand{\M}{\mathcal{M}}
\renewcommand{\L}{\mathcal{L}}
\renewcommand{\r}{\hat{\rho}}
\newcommand{\s}{\hat{\sigma}}

\allowdisplaybreaks[1] 

\begin{document}

\title{The geometry of spacetime from quantum measurements}

\author{T. Rick Perche}
\email{trickperche@perimeterinstitute.ca}
\affiliation{Department of Applied Mathematics, University of Waterloo, Waterloo, Ontario, N2L 3G1, Canada}
\affiliation{Perimeter Institute for Theoretical Physics, Waterloo, Ontario, N2L 2Y5, Canada}
\affiliation{Institute for Quantum Computing, University of Waterloo, Waterloo, Ontario, N2L 3G1, Canada}

\author{Eduardo Mart\'{i}n-Mart\'{i}nez}
\email{emartinmartinez@uwaterloo.ca}

\affiliation{Department of Applied Mathematics, University of Waterloo, Waterloo, Ontario, N2L 3G1, Canada}
\affiliation{Perimeter Institute for Theoretical Physics, Waterloo, Ontario, N2L 2Y5, Canada}
\affiliation{Institute for Quantum Computing, University of Waterloo, Waterloo, Ontario, N2L 3G1, Canada}

\begin{abstract}
    We provide a setup by which one can recover the geometry of spacetime from local measurements of quantum particle detectors coupled to a quantum field. Concretely, we show how one can  recover the field's correlation function from measurements on the detectors. Then, we are able to recover the invariant spacetime interval from the measurement outcomes, and hence reconstruct a notion of spacetime metric. This suggest that quantum particle detectors are the experimentally accessible devices that could replace the classical `rulers' and `clocks' of general relativity.
\end{abstract}

\maketitle

\section{Introduction}

    A full theory of quantum gravity is arguably one of the greatest challenges of modern theoretical physics. Ever since general relativity was first proposed, different approaches  have been attempted without no definite success to date. One of the obstacles for a full theory of quantum gravity lies in the fact that the current notions of time and space are defined in terms of the spacetime metric, which cannot be consistently quantized by usual means. 
    
    
    
    It is commonly accepted that at the Planck scale, no  experimentally verified theory provides a framework for measuring space or time. At these scales the theory of general relativity can be argued not to be valid anymore since there is no notion of ruler or clock. Thus, any method of introducing the notion of spacetime distance must rely on the only theory suitable on these scales: quantum field theory (QFT). 
    
    In recent pioneering work by Kempf et al.~\cite{achim,achimQGInfoCMB,achim2} it was shown that the notion of distance could be rephrased in terms of propagators of quantum fields. This rephrasing of distance and time intervals in terms of QFTs then allows one to write the observables of general relativity in terms of measurable quantities of quantum fields, and in the authors' opinion represents great progress towards our understanding of the relationship between these two theories.
    
    In this work we provide a physically realizable way of recovering the geometry of spacetime by means of local measurements of quantum field theories. The tools we consider for locally probing quantum fields are  the so-called particle detector models, such as the Unruh-DeWitt (UDW) detector~\cite{Unruh1976,Unruh-Wald,Takagi}. These tools have been successfully employed in the study of plenty of phenomena in quantum field theory, such as the Unruh effect~\mbox{\cite{Unruh1976,Unruh-Wald,Takagi,matsasUnruh,mine}} and Hawking radiation~\cite{HawkingRadiation,Sciama1977}.  Nowadays, many protocols of relativistic quantum information have been developed through the use of particle detector models, such as entanglement harvesting~\cite{Reznik1,reznik2,Valentini1991,RalphOlson1,Farming,ClicheKempfD,topology,Pozas2016,Sachs1,sachs2018entanglement}, quantum and classical communication~\cite{Benincasa_2014,Landulfo,Katja,Jonsson2,Jonsson3,Jonsson4,Shockwaves,Simidzija_2020}  and quantum energy teleportation~\cite{teleportation,HottaEntanglement,HottaDistance,nichoTeleport} to cite some. Moreover, particle detector models have a natural connection with experimental setups, being able model the interaction of atoms with light~\cite{eduardo,Jonsson1,Nicho1,richard} and nucleons with neutrinos~\cite{neutrinos,antiparticles} in more accurate ways than the typical models employed in quantum optics (e.g. Jaynes-Cummings) and nuclear physics\textcolor{black}{. Particle detectors have also} 
    proven to be respectful with fundamental principles of relativity theory such as causality~\cite{eduCausality,PipoFTL} and general covariance~\cite{us,us2}.
    
    Although particle detector models can be used to obtain local information about quantum fields, their  response is always a function of the $n$-point point  correlators of the  field, and not only of its propagators.  This adds an extra difficulty to directly apply the results  in~\cite{achim,achim2} to the response of detectors. To address this, in this manuscript we show that the metric can also be rebuilt from the two-point function of quantum fields that detectors can directly access~\cite{pipo}. This is due to the fact that the short distance behaviour of the field correlations in any physical state is only a function of the proper distance between events. Thus, it is possible to appropriately manipulate 
    the two-point function in order to recover the infinitesimal line element, or, equivalently, the spacetime metric.
    
    More concretely, we will extend the results in~\cite{pipo} about explicitly measuring the field Wightman function along the worldline of a particle detector to more general scenarios. {Specifically, we will show how to measure field} correlators at {timelike and spacelike} separated events with pairs of particle detectors. This provides an experimentally accessible protocol to probe the two point function of the field for any two close enough events. Then we use these results to provide a protocol with which the spacetime metric can be explicitly recovered from measurements of probes that couple to a quantum field. Moreover, we explicitly show that the short distance behaviour of the two-point function of a free real scalar quantum field is independent of any of its specific properties, such as the field mass or \emph{the field state}, allowing for the measurement of spacetime separations through the infinitesimal notion of distance provided by the correlation function of quantum fields captured by particle detectors. Finally, we also illustrate how the protocol works and the metric is recovered from measurements of the field in several example scenarios in flat and curved spacetimes.
    
    This work is organized as follows. In Section \ref{sec:geometry} we show how to recover the spacetime metric in terms of the quantum field's two-point function regardless of the field state. In Section \ref{sec:detectors} we discuss the UDW model and the setup used to locally probe the quantum field. We apply this setup to quantum field theories in curved spacetimes and show how to locally recover the metric and spacetime curvature in terms of the correlation functions between different UDW detectors. In Section \ref{sec:examples} we consider multiple examples and show explicitly how particle detectors can be used to recover the metric of usual spacetimes, such as Minkowski spacetime, hyperbolic static Robertson-Walker and deSitter spacetimes. We also consider with both pointlike and smeared particle detectors probing different field states. The conclusions of our work can be found in Section \ref{sec:conclusions}.

    \section{The spacetime geometry in terms of the two-point function of quantum fields}\label{sec:geometry}
    
        \textcolor{black}{This section will be devoted to establishing the relationship between the spacetime geometry and the field correlators as shown in~\cite{achim,achimQGInfoCMB,achim2}. In these references} it was shown that it is possible to recover the spacetime metric in terms of the Feynman propagator (time-ordered vacuum two-point function) for a scalar field defined in a curved spacetime. Here we write the results of~\cite{achim} in an explicitly coordinate independent formulation. We then use this to extend some of the results of \cite{achim,achimQGInfoCMB,achim2} showing that it is also possible to recover the metric by means of the (not time-ordered) two point function for \emph{any} field state. We will highlight that the reason why  one can recover the metric from both the Wightman function and the Feynman propagator is the specific dependence of these on the spacetime interval between events, something that \textcolor{black}{is necessary for the fulfillment of the Hadamard condition~\cite{fullingHadamard,fullingHadamard2,kayWald,equivalenceHadamard,fewsterNecessityHadamard}}.
        
        As a first step, before going through how to recover the metric in terms of correlators and propagators, we briefly review the basics of Synge's world function and Klein-Gordon quantum fields in curved spacetimes.
            
        \subsection{Synge's world function}\label{sub:Synge}
        
            In this subsection we briefly discuss the properties of Synge's world function, $\sigma(\mf x,\mf x')$. In a few words, Synge's function takes as input two sufficiently close events in spacetime and returns one half of the squared geodesic distance between them. That is, if $\mathcal{M}$ is a $D=n+1$ dimensional spacetime, with $n\geq 2$, and $\mf x,\mf x'\in\mathcal{M}$ are two events that can be connected by a unique geodesic $\gamma:[u_1,u_2]\rightarrow \mathcal{M}$, Synge's world function can be written as
            \begin{equation}
                \sigma(\mf x,\mf x') = \frac{1}{2}(u_2-u_1) \int_{u_1}^{u_2} \dd u \: g_{\mu\nu}(\gamma(u))\dv{\gamma^\mu}{u}\dv{\gamma^\nu}{u}.
            \end{equation}
            Or, in the particular case where $\gamma$ is parametrized by its proper time/length parameter $s\in[0,r]$ with $\gamma(0) = \mf x$ and $\gamma(r) = \mf x'$, we can write the simpler expression
            \begin{equation}
                \sigma(\mf x,\mf x') = \frac{r^2}{2}.
            \end{equation}
            The domain of $\sigma$ is then the subset of $\mathcal{M}\times \mathcal{M}$ consisting of points that can be connected by a unique geodesic. It is important to notice that given $\mf x\in\mathcal{M}$, there always exists its normal neighbourhood: the largest open set such that every point within it can be connected to $\mf x$ by a unique geodesic. This implies that $\sigma(\mf x,\mf x')$ is always locally defined around each point.

            Synge's world function is particularly useful to talk about expansions of tensors around a spacetime point. $\sigma(\mf x,\mf x')$ can be differentiated with respect to each of its arguments, and its derivatives can be used to extend the notion of separation vector to curved spacetimes. In fact, it is possible to show that the derivatives of $\sigma(\mf x,\mf x')$ are related to the initial tangent vector (i.e., at $\mf x$) of the geodesics that connect $\mf x$ and $\mf x'$.
            
           It is usual to denote differentiation with respect to the first argument with  no prime and  differentiation with respect to the second argument with primes. Namely, we denote
            \begin{align}\label{eq:indices}
                \nabla_{\mu} \sigma = \pdv{}{x^\mu} \sigma(\mf x,\mf x')&&& \nabla_{\mu'}\sigma = \pdv{}{x'{}^\mu}\sigma(\mf x,\mf x').
            \end{align}
            Notice that both elements above can be seen as components of a tensor of rank $(0,1)$ that depends on two spacetime points. However, each of the tensors in Eq. \eqref{eq:indices} lie in different tangent spaces. That is, $\nabla_\mu\sigma\in T_{\mf x}\mathcal{M}$, while  $\nabla_{\mu'}\sigma\in T_{\mf x'}\mathcal{M}$. Moreover, it is usual to denote derivatives of the world function by adding an index to $\sigma$ instead of explicitly writing the symbol $\nabla$. We will use this notation in the manuscript and simply denote
            \begin{align}
                \sigma_{\alpha'} \equiv \nabla_{\alpha'}\sigma &&& \sigma_{\beta\alpha}\equiv\nabla_{\alpha}\nabla_{\beta}\sigma,
            \end{align}
            and analogous expressions for higher derivatives. For a detailed review of Synge's World function, and the usual conventions, we refer the reader to \cite{poisson}.
            
            As we mentioned above, Synge's world function can be used to obtain the tangent vector to the geodesics that connect nearby points, thus generalizing the concept of separation vector between them. More specifically, the vector $\sigma^\mu(\mf x,\mf x')\in T_{\mf x}\mathcal{M}$ is the tangent vector to the geodesic that connects $\mf x$ and $\mf x'$ such that its squared norm corresponds to the squared spacetime interval between these events. That is,
            \begin{equation}
                \sigma = \frac{1}{2}\sigma^\mu \sigma_\mu.
            \end{equation}
            The vector $\sigma^\mu\in T_{\mf x}\mathcal{M}$ can also be used to build Riemann normal coordinates around the point $\mf x$. In fact, $\sigma^\mu(\mf x,\mf x')$ corresponds to the ``position vector'' of the point $\mf x'$ in Riemann normal coordinates. Conversely, $\sigma^{\mu'}$ is the tangent vector (at $\mf x'$) to the geodesic that starts at $\mf x'$ and reaches $\mf x$, with squared norm equal to $2\sigma(\mf x,\mf x')$. Synge's world function then allows us to write expressions in terms of Riemann normal coordinates around an event in completely covariant language, without explicitly relying on coordinate systems.

            The notion of coincidence limit is of particular relevance in this manuscript. The coincidence limit is the name given to the limit of a bitensor that depends on $\mf x$ and $\mf x'$ when $\mf x'\rightarrow \mf x$. In essence, this limit brings a tensor defined at different points to a tensor defined at $\mf x$. Following the usual notation in the literature, we denote the coincidence limit of a bitensor by brackets. For instance,
            \begin{equation}\label{eq:coinSynge}
                [\sigma_{\alpha\beta'}] \coloneqq \lim_{\mf x'\rightarrow \mf x} \sigma_{\alpha\beta'}.
            \end{equation}
            Notice that  Eq.~\eqref{eq:coinSynge} {is the coincidence limit of an object that is} a (0,1) tensor at $\mf x$ and a type (0,1) tensor at $\mf x'$. {The limit yields a} (0,2) tensor at $\mf x$. The coincidence limits of Synge's world function will allow us to recover the spacetime metric from the propagators and two-point functions of a quantum field. We recall the following coincidence limit identities for Synge's world function~\cite{poisson}:
            \begin{align}
                &[\sigma_\alpha] = [\sigma_{\alpha'}] = 0\\
                &[\sigma_{\alpha\beta}] = [\sigma_{\alpha'\beta'}] = g_{\alpha\beta},\label{eq:rec1}\\
                &[\sigma_{\alpha\beta'}] = [\sigma_{\alpha'\beta}] = -g_{\alpha \beta},\label{eq:rec2}\\
                &[\sigma_{\alpha\beta\gamma\delta}] =[\sigma_{\alpha\beta\gamma'\delta'}] =[\sigma_{\alpha'\beta'\gamma'\delta'}]= \frac{2}{3}R_{\alpha(\gamma\delta)\beta},\label{eq:rec3}\\
                &[\sigma_{\alpha\beta'\gamma'\delta'}] = \frac{2}{3}R_{\alpha(\beta\gamma)\delta}.\label{eq:rec4}
            \end{align}
            Notice that since the coincidence limit is always a tensor at the point $\mf x$, there are no primed indices on the rightmost term of the equations above. As we can see in expressions \eqref{eq:rec1} and \eqref{eq:rec2}, it is then possible to recover the spacetime metric at a point $\mf x$ by taking limits of derivatives of Synge's world function. 
            
            {\color{black}
            
            We end this section with two other examples of bitensors: the parallel propagator $g^{\alpha'}{}_{\beta}(\mf x,\mf x')$ and the Van-Vleck determinant, $\Delta(\mf x,\mf x')$, defined in the same domain as $\sigma(\mf x,\mf x')$. The parallel propagator is defined as the bitensor that maps a vector $v\in T_{\mf x}\mathcal{M}$ to its parallel transport at $T_{\mf x'}\mathcal{M}$ along the unique geodesic that connects $\mf x$ with $\mf x'$. It is intuitive that its coincidence limit reads $[g^{\alpha'}{}_{\beta}] = \delta^\alpha_\beta$. The parallel propagator is related to Synge's world function via $\sigma^{\alpha'} = g^{\alpha'}{}_\beta\sigma^\beta$, which can be seen from the fact that tangent vectors to geodesics are parallel transported. The Van-Vleck determinant $\Delta(\mf x,\mf x')$ is defined as
            \begin{equation}
                \Delta(\mf x,\mf x') = \det(-g^{\alpha'}{}_{\alpha}(\mf x,\mf x')\sigma^{\alpha}{}_{\beta'}(\mf x,\mf x')),
            \end{equation}
            and it naturally arises in expressions for Green's functions for different field equations of motions, as it is the solution to the differential equation $\nabla_\alpha(\Delta \sigma^\alpha) = 4\Delta$. {Additionally, it is important to notice that} as $\mf x'\rightarrow \mf x$, the Van-Vleck determinant behaves as $\Delta = 1 + \mathcal{O}(\sigma^2)$.}

        \subsection{The free real scalar field in curved spacetimes}\label{sub:KG}
        
            In this subsection we will briefly review the quantization of a real scalar quantum field in curved spacetimes. We consider a $D = n+1$ dimensional spacetime $\mathcal{M}$ with a real scalar field ${\phi}(\mf x)$ associated with the Lagrangian
            \begin{equation}
                \mathcal{L} = -\frac{1}{2}\nabla_\mu\phi \nabla^\mu\phi - \frac{1}{2}\left(m^2 +\xi R\right)\phi^2,\label{eq:lag}
            \end{equation}
            where $m$ is the field's mass, $R$ is the Ricci scalar and $\xi$ determines the coupling of the field to curvature. Two specific values of $\xi$ that are interesting are the conformally coupled case with $\xi = (D-2)/4(D-1)$ and the minimally coupled case with $\xi = 0$. The equation of motion that arises from the extremization of the action associated with the Lagrangian in \eqref{eq:lag} is the Klein-Gordon equation,
            \begin{equation}\label{eq:KG}
                (\nabla_\mu\nabla^\mu -m^2-\xi R)\phi = 0.
            \end{equation}
            
            We quantize the field by choosing a complete set of solutions to the equation that is orthonormal with respect to the (non-positive) inner product defined by the continuity equation. We denote this set of \textit{modes} by $\{u_{\bm k}(\mf x),u_{\bm k}^*(\mf x)\}$, and imposing canonical commutation relations so that the field operator can be written as
            \begin{equation}\label{eq:modeExp}
                \hat{\phi}(\mf x) = \int \dd^n \bm k\left(u_{\bm k}(\mf x) \hat{a}_{\bm k} + u_{\bm k}^*(\mf x) \hat{a}^\dagger_{\bm k}\right),
            \end{equation}
            where $\hat{a}_{\bm k}$ are the annihilation operators associated with this mode decomposition.    It is then possible to construct a Hilbert space representation for the QFT, from the vacuum state $\ket{0}$, assumed to be the only state such that $\hat{a}_{\bm k}\ket{0} = 0$. The Fock space associated with this mode decomposition is then constructed by repeated applications of the creation operators $\hat{a}^\dagger_{\bm k}$ to $\ket{0}$. The creation and annihilation operators satisfy the canonical commutation relations,
            \begin{equation}
                \big[\hat{a}^{\vphantom{\dagger}}_{\bm k},\hat{a}^\dagger_{\bm k'}\big] = \delta^{(n)}(\bm k - \bm k').
            \end{equation}
            
            In QFT there are two scalar distributions that are particularly relevant: the Feynman propagator, $G_F(\mf x,\mf x')$, and the vacuum Wightman function, $W(\mf x,\mf x')$. These are defined by the following expressions
            \begin{align}
                G_F(\mf x,\mf x') &= \bra{0}\!\mathcal{T}\hat{\phi}(\mf x)\hat{\phi}(\mf x')\!\ket{0},\\
                W(\mf x,\mf x') &= \bra{0}\!\hat{\phi}(\mf x)\hat{\phi}(\mf x')\!\ket{0},
            \end{align}
            where $\mathcal{T}$ is the time ordering operation. {\color{black} As distributions, $G_F(\mf x,\mf x')$ and $W(\mf x,\mf x')$ act on sufficiently regular functions $f$ and $g$ according to 
            \begin{align}
                G_F(f,g) &= \int_{\M\times\M}\!\!\!\!\!\!\dd V \,\dd V' f(\mf x) g(\mf x') G_F(\mf x,\mf x'),\label{eq:GF}\\
                W(f,g) &= \int_{\M\times\M}\!\!\!\!\!\!\dd V \,\dd V' f(\mf x) g(\mf x') W(\mf x,\mf x'),\label{eq:W}
            \end{align}
            where $\dd V$ is the invariant spacetime volume element. However, the integrals above are singular when \mbox{$\sigma(\mf x,\mf x')\rightarrow 0$}, so that a regulator $i \epsilon$ must be added which indicates the contour in the complex plane will be taken to solve the integrals \eqref{eq:GF} and \eqref{eq:W}. After computing the integrals the limit $\epsilon \rightarrow 0$ can be taken, yielding a finite result~\cite{kayWald}. 
            
            The Wightman function is particularly important in order to define which quantization schemes can be associated to physical vacuum states. It can be shown that the expected value of the stress energy momentum tensor in $\ket{0}$ can only be renormalized if the vacuum state satisfies the so called Hadamard condition~\cite{fullingHadamard,fullingHadamard2,kayWald,equivalenceHadamard,fewsterNecessityHadamard}. In essence, $\bra{0}\!\hat{T}_{\mu\nu}\!\ket{0}$ can only be defined if the Wightman function can be written as
            \begin{equation}\label{eq:Hadamard}
                W(\mf x,\mf x') = \frac{1}{8\pi^2} \frac{\Delta^{1/2}(\mf x,\mf x')}{\sigma(\mf x,\mf x')} + v(\mf x,\mf x') \text{ln}|\sigma(\mf x,\mf x')|+h(\mf x,\mf x'),
            \end{equation}
            where $\Delta(\mf x,\mf x')$ is the Van-Vleck determinant, $\sigma(\mf x,\mf x')$ is Synge's world function, $h(\mf x,\mf x')$ is a regular function \tb{that contains the state dependence} and $v(\mf x,\mf x')$ can be written as a power series in $\sigma(\mf x,\mf x')$, whose coefficients are determined by the Hadamard recursion relations~\cite{DeWittExpansion,poisson}. In this manuscript we will assume that the vacuum defined by the mode expansion in Eq. \eqref{eq:modeExp} is a Hadamard state.}
            
            The Feynman propagator can also be related to the Wightman function by
            \begin{align}
                G_F(\mf x,\mf x') &= W(\mf x,\mf x')\theta(t-t') + W(\mf x',\mf x)\theta(t'-t)\nonumber\\
                &= \Re[ W(\mf x,\mf x')] + \ii\varepsilon(t-t')\Im[ W(\mf x,\mf x')],
            \end{align}
            where $\theta(t)$ denotes the Heaviside step function and \mbox{$\varepsilon(t) = \theta(t)-\theta(-t)$} denotes the sign function. In particular, due to the fact that the imaginary part of the Wightman function goes to zero in the coincidence limit, it is possible to formally write
            \begin{equation}\label{eq:WD}
                \lim_{\mf x' \rightarrow \mf x}G_F(\mf x,\mf x') = \lim_{\mf x' \rightarrow \mf x}W(\mf x,\mf x'),
            \end{equation}
            where the equality has to be understood in a distributional sense.

        \subsection{The spacetime metric in terms of the Wightman function}\label{sub:metricW}

            We start this subsection by reformulating the results of~\cite{achim} in an explicitly covariant way in terms of Synge's world function and coincidence limits. In \cite{achim} it was shown that in a $D = n+1$ dimensional spacetime, the metric can be obtained from the Feynman propagator of Klein-Gordon's equation via the expression
            \begin{equation}
                g_{\mu\nu} = -\frac{1}{2}\left(\frac{\Gamma\left(\frac{D}{2}-1\right)}{4 \pi^{D / 2}}\right)^{\frac{2}{D-2}}\!\!\!\!\!\!\!\lim_{\mf x'\rightarrow \mf x} \pdv{}{x^\mu}\pdv{}{x'^\nu}\left(G_F(\mf x,\mf x')\right)^{\frac{2}{2-D}},
            \end{equation}
            where $G_F(\mf x,\mf x')$ is the Feynman propagator associated with a given quantization framework. This result was established by studying the short distance behaviour of the Klein-Gordon equation with a Dirac-delta source. In \cite{achim}, the equation for the propagator was solved using Riemann normal coordinates around a point.
            
            It is possible to rewrite Eq. (C21) in~\cite{achim} for the short-separation behaviour of the Feynman propagator in terms of Synge's world function:
            \begin{equation}\label{eq:achim}
                G_F(\mf x,\mf x') = G_{0}(\sigma(\mf x,\mf x'))(1 + f(\mf x,\mf x')),
            \end{equation}
            where $G_0(\sigma)$ is the massless Minkowski spacetime Feynman propagator as a function of the half squared spacetime interval between events, $\sigma$, and $f(\mf x,\mf x')\rightarrow 0$ as $\mf x' \rightarrow \mf x$. $G_{0}(\sigma)$ can be explicitly computed, and its short distance behaviour is given by
            \begin{equation}\label{eq:delta0}
                G_{0}(\sigma) =  \frac{\Gamma\left(\frac{D}{2}-1\right)}{2 (2\pi)^{D / 2}} \frac{1}{\sigma^{\frac{D-2}{2}}},
            \end{equation}
            which corresponds to the Wightman function of a massless scalar field in Minkowski spacetime. The fact that the two coincide close to the coincidence limit is not surprising as argued in Eq.~\eqref{eq:WD}.
            

            Hence, by taking the $2/(2-D)$th power of Eq.~\eqref{eq:achim}, one obtains the proportionality
            \begin{align}\label{eq:propto}
                (G_{F}(\mf x,\mf x'))^{\frac{2}{2-D}} \propto \sigma(\mf x,\mf x'),
            \end{align}
            where the remaining terms are negligible in the short distance limit. If one now wishes to recover the spacetime metric, it is enough to compute the derivatives of Synge's world function with respect to its different arguments together with the coincidence limits in Eq. \eqref{eq:rec2}. In fact, by differentiating \eqref{eq:propto} with respect to $\mf x$ and $\mf x'$, one obtains
            \begin{equation}
                \partial_{\mu}\partial_{\nu'}(G_{F}(\mf x,\mf x'))^{\frac{2}{2-D}} \approx \left(\frac{\Gamma\left(\frac{D}{2}-1\right)}{2 (2\pi)^{D / 2}}\right)^{\frac{2}{2-D}}\sigma_{\mu\nu'},
            \end{equation}
            where the ``$\approx$'' symbol denotes that the expression above is only valid when $\mf x'\rightarrow \mf x$. By taking the coincidence limit on both sides and isolating the $[\sigma_{\mu\nu'}]$ term, we then have
            \begin{equation}\label{eq:mainAchim}
                -g_{\mu\nu} =\! [\sigma_{\mu\nu'}]\! = \frac{1}{2}\! \left(\frac{\Gamma\left(\frac{D}{2}-1\right)}{4 \pi^{D / 2}}\right)^{\frac{2}{D-2}}\!\!\!\!\!\!\![\partial_{\mu}\partial_{\nu'}(G_{F}(\mf x,\mf x'))^{\frac{2}{2-D}}],
            \end{equation}
            which establishes the result of Eq. (6) in \cite{achim} in an explicitly covariant manner.
            
            We now shift focus to the vacuum Wightman function of a real scalar quantum field. In essence, in order to write the metric in terms of the Wightman function, we notice that---same as the Feynman propagator---the Wightman function $W(\mf x,\mf x')$ for a massive field can  be written in terms of the massless (vacuum) Wightman function in Minkowski spacetime ($G_{0}$ in Eq.~\eqref{eq:delta0}) added to corrections that become negligible in the short distance limit. This implies that Eq. \eqref{eq:mainAchim} still holds when one replaces the Feynman propagator $G_{F}(\mf x,\mf x')$ by the vacuum Wightman function $W(\mf x,\mf x')$. This result can also be obtained from Eq. \eqref{eq:WD}, where we see that the real part of $G_F(\mf x,\mf x')$ and of $W(\mf x,\mf x')$ is the same. Given that the limit yields a real function (the components of the metric $g_{\mu\nu}$ are real), we obtain the desired result. Namely, we can write the metric as the following coincidence limit of the vacuum Wightman function,
            \begin{equation}\label{mine0}
                g_{\mu\nu} =  -\frac{1}{2} \left(\frac{\Gamma\left(\frac{D}{2}-1\right)}{4 \pi^{D / 2}}\right)^{\frac{2}{D-2}}[\partial_{\mu}\partial_{\nu'}(W(\mf x,\mf x'))^{\frac{2}{2-D}}].
            \end{equation}
            
            We now discuss whether the assumption that the field is in the vacuum state is necessary for this protocol. In Appendix \ref{app:stateind}, we show that the Wightman function on \emph{any} normalized field state $\hat{\rho}_{\phi}$ can always be written as
            \begin{align}
                W_{\rho_{\phi}}(\mf x,\mf x')&\coloneqq \ev{\hat{\phi}(\mf x)\hat{\phi}(\mf x')}_{\hat{\rho}} \nonumber\\&= W(\mf x,\mf x')+ \sum_{m=0}^\infty F_m(\mf x) G^*_m(\mf x')+\text{H.c.},
            \end{align}
            for a given set of functions $F_m$ and $G_m$. We also show that these functions are regular in the limit $\mf x'\rightarrow \mf x$. In particular, this implies that they are negligible compared to the (singular) vacuum Wightman function in this limit. \textcolor{black}{ As discussed in Subsection \ref{sub:KG},} any Hadamard state of a quantum field theory will \textcolor{black}{be such that the Wightman function can be written as in Eq. \eqref{eq:Hadamard}. As we show in Appendix \ref{app:stateind}, the singular part of the Wightman function of any state in a given quantization scheme is the same as that of the vacuum. That is, using the quantization scheme discussed in Subsection \ref{sub:KG} and under the assumption that $\ket{0}$ is a Hadamard state, we can write}
            \begin{equation}\label{mine}
                g_{\mu\nu} =  -\frac{1}{2} \left(\frac{\Gamma\left(\frac{D}{2}-1\right)}{4 \pi^{D / 2}}\right)^{\frac{2}{D-2}}[\partial_{\mu}\partial_{\nu'}(W_{\rho_{\phi}}(\mf x,\mf x'))^{\frac{2}{2-D}}]
            \end{equation}
            for any normalized field state $\hat{\rho}_{\phi}$. In particular, this implies that no matter the state of the field, it is possible to recover the metric by means of the limit of Eq. \eqref{mine}. {This will be key in order for us to} recover the geometry of spacetime {from local measurements} of the quantum fields.
            
            
            It is worth noting that it is not straightforward to try to recover the spacetime curvature directly via further differentiation from the Wightman function  by using Eqs. \eqref{eq:rec3} and \eqref{eq:rec4}. This is so because repeated differentiations of a scalar bitensor with respect to the same point require covariant derivatives. Thus, we would need to have prior knowledge of the connection in order to perform these computations. However, it is true that the spacetime curvature can be written as
            \begin{equation}
                R_{\alpha(\gamma\delta)\beta}= \frac{3}{4} \!\left(\frac{\Gamma\left(\frac{D}{2}-1\right)}{4 \pi^{D / 2}}\right)^{\frac{2}{D-2}}\!\!\!\!\!\!\![\nabla_{\alpha}\partial_{\beta}\nabla_{\gamma'}\partial_{\delta'}(W(\mf x,\mf x'))^{\frac{2}{2-D}}],
            \end{equation}
            which comes from Eq. \eqref{eq:rec3}. The equivalent computations that come from  Eqs. \eqref{eq:rec3} and \eqref{eq:rec4} are also valid, but again, require prior knowledge of the connection to compute the right hand side. The same is true if one tries to differentiate $(W(\mf x,\mf x'))^{\frac{2}{2-D}}$ twice with respect to the same argument and employ Eq. \eqref{eq:rec1} to recover the metric: we would need to know the Christoffel symbols in order to obtain the metric through this procedure.

\section{Locally probing a quantum field}\label{sec:detectors}

    In this section we will consider multiple particle detectors probing a real scalar quantum field so that the information they obtain can be used to extract the spacetime metric. In \cite{pipo} it was shown how to use one particle detector to recover the two-point function of a real scalar field between two timelike separated events {using} detector measurements. Here we will review this concept and extend it by showing how to use two detectors to recover the two-point function of the quantum field for spacelike separated events in terms of the correlations acquired by the detectors.
    
    \subsection{The Setup}\label{sub:setup}
    
    Consider a real scalar quantum field $\hat{\phi}(\mf x)$ in an $(n+1)$-dimensional globally hyperbolic spacetime $\mathcal{M}$. Let $\Sigma_t$ be a foliation of $\mathcal{M}$ by a family of Cauchy surfaces, where $t$ denotes a future oriented timelike coordinate. We will consider a set of particle detectors that undergo timelike trajectories $\mf z_i(\tau_i)$ in $\mathcal{M}$, where $\tau_i$ is denotes their respective proper times. We use the convention that at $\Sigma_0$, the $\tau_i$ parameters also take the value $0$, in order to simplify future computations.
    
    We assume all detectors to be two-level quantum systems with the same proper energy gap $\Omega$. Their free Hamiltonians (generating translations with respect to their respective proper time $\tau_i$) are given by
    \begin{equation}
        \hat{H}_i = \Omega \hat{\sigma}_i^+ \hat{\sigma}_i^-,
    \end{equation}
    where $\hat{\sigma}_i^\pm$ are the raising and lowering operators for each detector two-level system. We further assume the detectors to couple with the same coupling strength {$\lambda$} to the field, according to the following scalar interaction Hamiltonian weight\footnote{The Hamiltonian weight is related to the Hamiltonian density as follows $\mathfrak{h}_I(\mf x)$ by $\mathfrak{h}_I(\mf x) = \sqrt{-g} \hat{h}_I(\mf x)$.}:
    \begin{equation}
        \hat{h}_i(\mf x) = \lambda \Lambda_i(\mf x)\hat{\mu}_i(\tau_i)\hat{\phi}(\mf x),
    \end{equation}
    where $\hat{\mu}_i(\tau_i) = e^{\ii\Omega \tau_i}\sigma^+_i + e^{-\ii\Omega \tau_i}\sigma^-_i$ is the monopole moment of the $i$th probe, $\Lambda_i(\mf x)$ is a function that controls the spacetime localization of the interaction of the probes with the field.
    
    
    We will assume the probes to start in their ground states $\ket{g_i}\!\!\bra{g_i} = \hat{\sigma}_i^- \hat{\sigma}_i^+$, and the field to be in an arbitrary state $\hat{\rho}_{\phi}$ so that the initial state of the full system will be given by
    \begin{equation}
        \hat{\rho}_0 = \bigotimes_i \hat{\sigma}_i^- \hat{\sigma}_i^+\otimes\hat{\rho}_{\phi}.
    \end{equation}
    We are interested in computing the final state of the system, given by
    \begin{equation}
        \hat{\rho} = \hat{U} \hat{\rho}_0 \hat{U}^\dagger,
    \end{equation}
    where $\hat{U}$ is the time evolution operator,
    \begin{equation}
        \hat{U} = \mathcal{T}\exp\left(-\ii\int \dd V \: \hat{h}_I(\mf x)\right), \quad \hat{h}_I(\mf x) \equiv \sum_i \hat{h}_i(\mf x),
    \end{equation}
    where $\dd V$ is the invariant spacetime volume element. To proceed, we employ a perturbative approach, by means of the Dyson expansion, $\hat{U} = \openone + \hat{U}^{(1)} + \hat{U}^{(2)} + ...$, where we assume each of the $\hat{U}^{(k)}$ terms to be of order $k$ in the coupling constants $\lambda$. The $\hat{U}^{(k)}$ terms are explicitly given by
    \begin{equation}
        \hat{U}^{(k)} = \frac{(-\ii)^k}{k!}\int... \int \dd V_1... \dd V_k \mathcal{T}\hat{h}_I(\mf x_1)... h_I(\mf x_k),
    \end{equation}
    where $\mathcal{T}$ denotes the time ordering operation. With this, the final state of the system after the interaction will be given by
    \begin{align}
        \r = \r_0& + \sum_{k=1}^\infty \r^{(k)},
    \end{align}
    where
    \begin{align}
       \r^{(k)} = \sum_{l=0}^k\hat{U}^{(l)} \r_0 \hat{U}^{(k-l)\dagger}.
    \end{align} 
    For simplicity, we will work under the assumption that the initial state of the field $\hat\rho_\phi$ is a zero-mean Gaussian state (e.g., vacuum, squeezed vacuum, thermal states, etc), so that the two-point function of the system will be enough to fully describe it.
    
    The next step to compute the final state of the detectors  after the interaction, $\r_{\textsc{d}}$, {is to evaluate} the partial trace over the field's degrees of freedom, \mbox{$\r_{\textsc{d}} = \tr_\phi (\r)$}. 
    
    The state of the field-detectors system after the interaction, to second order in perturbation theory, is given by $\r= \r_0 + \r^{(2)} +\mathcal{O}(\lambda^4)$, where the second order correction is
    \begin{align}
        \r^{(2)} = -\lambda^2\sum_{ik}&\int \dd V \dd V'\Lambda(\mf x)\Lambda(\mf x')\\
        &\Big(\theta(t-t')\hat{\mu}_i(\tau_i)\hat{\mu}_k(\tau_k')\hat{\rho}_{{\textsc{d}}}^{(0)}\hat{\phi}(\mf x)\hat{\phi}(\mf x') \hat{\rho} \nonumber\\&- \hat{\mu}_i(\tau_i)\hat{\rho}_{\textsc{d}}^{(0)}\hat{\mu}_k(\tau_k')\hat{\phi}(\mf x)\hat{\rho}\hat{\phi}(\mf x')\nonumber\\
        &+ \theta(t'-t)\hat{\rho}_{\textsc{d}}^{(0)}\hat{\mu}_i(\tau_i)\hat{\mu}_k(\tau_k') \hat{\rho}\hat{\phi}(\mf x)\hat{\phi}(\mf x')\Big).\nonumber
    \end{align}
    partial tracing over the field's degree of freedom, we obtain
    \begin{align}
        &\r^{(2)}_\textsc{d} = -\lambda^2\sum_{ik}\int \dd V \dd V'\Lambda_i(\mf x)\Lambda_k(\mf x')\ev{\hat{\phi}(\mf x)\hat{\phi}(\mf x')}\nonumber\\
        &\Big(\theta(t-t')\hat{\mu}_i(\tau_i)\hat{\mu}_k(\tau_k')\hat{\rho}_{\textsc{d}}^{(0)} - \hat{\mu}_k(\tau_k')\hat{\rho}_{\textsc{d}}^{(0)}\hat{\mu}_n(\tau_n)\\&+ \theta(t'-t)\hat{\rho}_{\textsc{d}}^{(0)}\hat{\mu}_i(\tau_i)\hat{\mu}_k(\tau_k') \Big).
    \end{align}
    And we have
    \begin{align}
        &\hat{\mu}_i(\tau_i)\hat{\mu}_k(\tau_k')\hat{\rho}_{0} = \begin{cases}
            \bigotimes_{m} \s_m^-\s_m^+ e^{-\ii\Omega(\tau_i-\tau_i')}, i= k\\
            \bigotimes_{m\neq i,k} \s_m^- \s_m^+ \s_k^+ \s_i^+ e^{\ii\Omega(\tau_i+\tau_k')}
        \end{cases}\\
        &\hat{\rho}_{0}\hat{\mu}_i(\tau_i)\hat{\mu}_k(\tau_k') = \begin{cases}
            \bigotimes_{m} \s_m^-\s_m^+ e^{-\ii\Omega(\tau_i-\tau_i')}, i= k\\
            \bigotimes_{m\neq k,i}\s_m^-\s_m^+ \s_k^- \s_i^- e^{-\ii\Omega(\tau_i+\tau_k')}
        \end{cases}\\
        &\hat{\mu}_k(\tau_k')\hat{\rho}_0\hat{\mu}_i(\tau_i) = \begin{cases}
            \bigotimes_{m\neq i} \s_m^-\s_m^+  \s^+_i\s_i^- e^{-\ii\Omega(\tau_i-\tau_k')}, i=k\\
            \bigotimes_{m\neq k,i} \s_m^-\s_m^+ \s_i^+  \s_k^- e^{-\ii\Omega(\tau_i-\tau_k')}
        \end{cases}
    \end{align}
    With these, we obtain:
    \begin{align}\label{eq:rho2}
        &\r_\textsc{d}^{(2)} = \bigotimes_i \hat{\sigma}^-_i\hat{\sigma}^+_i + \lambda^2 \sum_i \bigotimes_{m\neq i} \hat{\sigma}^-_m \hat{\sigma}^+_m \comm{\hat{\sigma}^+_i}{\hat{\sigma}^-_i} \mathcal{L}_{ii} \\&\!\!+\!\lambda^2 \!\sum_{i\neq k} \bigotimes_{m\neq i,k}\!\! \hat{\sigma}^-_m \hat{\sigma}^+_m (\hat{\sigma}^-_k\hat{\sigma}^+_i \mathcal{L}_{ik}\!-\!\hat{\sigma}^+_k \hat{\sigma}^+_i\! \mathcal{N}_{ik}\!-\hat{\sigma}^-_k \hat{\sigma}^-_i \mathcal{N}_{ik}^*)\nonumber,
    \end{align}
    where 
    \begin{align}
        \mathcal{L}_{ik} &=\!\!\! \int\!\!\dd V \dd V'\Lambda_i(\mf x)\Lambda_k(\mf x') e^{-\ii\Omega(\tau_i-\tau_k')}\!\ev{\hat{\phi}(\mf x)\hat{\phi}(\mf x')},\label{eq:Lik}\\
        \mathcal{N}_{ik} \!&=\!\!\!\int \!\!\dd V \dd V'\Lambda_i(\mf x)\Lambda_k(\mf x') e^{\ii\Omega(\tau_i+\tau_k')}\theta(t-t')\!\ev{\hat{\phi}(\mf x)\hat{\phi}(\mf x')}\!.\nonumber
    \end{align}
    The terms associated to each individual detector are given by the diagonal terms
    \begin{equation}\label{eq:LiiFirst}
        \mathcal{L}_{ii} = \int \!\!\dd V \dd V'\Lambda_i(\mf x)\Lambda_i(\mf x') e^{-\ii\Omega(\tau_i-\tau_i')}\!\!\ev{\hat{\phi}(\mf x)\hat{\phi}(\mf x')}.
    \end{equation}
    These terms are {precisely} the excitation probability of each of the individual detectors. These probabilities were thoroughly studied in \cite{pipo}, where it was shown that it is possible to reconstruct the two-point function for the observable $\hat{\phi}(t,\bm x)$ for timelike separated events in terms of the transition probabilities of the detector and the proper time separation of the events. 
    
    \tb{We remark that if the interaction regions defined by $\Lambda_i(\mf x)$ are non-pointlike, all quantities in the second order expansion considered in this section are finite. Moreover, even when $\Lambda_i(\mf x)$ are singular Dirac delta functions, we have that the $\mathcal{L}_{ij}$ terms are finite for $i\neq j$, even though the individual excitation probabilities diverge. The reason for the divergence of the $\mathcal{L}_{ii}$ terms in this limit is that the integrals that define it sample the correlation function only at $\mf x = \mf x'$, and $\langle \hat{\phi}(\mf x) \hat{\phi}(\mf x')\rangle$ is divergent at this limit.}

\subsection{Measuring the two-point function of the quantum field}\label{sec:explicit}

    In this subsection, we will review how to extract the two-point function of a scalar quantum field extending the results~\cite{pipo} for timelike separation and generalizing these results to the case of spacelike separated events.

\subsubsection{Timelike separated events}\label{sub:timelike}
    
    As discussed above the two-point correlation function of a quantum field in timelike separated events can be recovered from the excitation probability of a single particle detector interacting with the field twice~\cite{pipo}. As mentioned in Subsection \ref{sub:setup}, this probability for the $i$th detector interacting with the field is given by the term $\mathcal{L}_{ii}$.
    
    The $\mathcal{L}_{ii}$ term depends on the spacetime localization of the interaction between the detector and the field. For simplicity, we assume a very fast switching modelled by a delta coupling\footnote{It is important to think of the delta coupling as a mathematical approximation for very fast switching. We remark that a delta coupling can introduce divergences in the model (that were studied in detail in~\cite{pipo,nogo,Petar}). However, these divergences can appear in the local terms associated with each interaction, and do not play any role in the correlations between detectors that we will use for the purpose of recovering the metric. If e.g., very sharp Gaussians are used instead of Dirac deltas we would obtain basically the same results for the correlations while keeping the system divergence-free at the price of complicating the calculations beyond the scope of the paper. For all purposes, one can think that we take any spacetime smearing function $\Lambda(\mf x)$ centered at $\mf x_i$ with unit integral and consider the family of spacetime smearings $\frac{1}{\eta^4}\Lambda(\mf x/\eta)$ in the limit $\eta \rightarrow 0$.} (see, e.g.,~\cite{Petar}) between the detector and field twice at two different times, at $\tau_i = \text{t}_1$ and at the latter time $\tau_i = \text{t}_2$. \tb{Although a true delta-coupling interaction is unphysical, it can be well modelled by interactions of sufficiently small systems that interact with the quantum field for times of the order of their light-crossing time. Nevertheless, it is important to mention that for  interactions that happen this fast are extremely difficult to be implemented in experimental setups.}
    
    \tb{The assumption of delta-coupling translates} into the following choice of spacetime smearing function    
    \begin{equation}\label{eq:timelikeLambda}
        \Lambda_i(\mf x) = \delta(\mf x - \mf z_i(\text{t}_1))/\sqrt{-g}+\delta(\mf x - \mf z_i(\text{t}_2))/\sqrt{-g}.
    \end{equation}
    In Appendix \ref{app:corrfunc} we show that this choice for $\Lambda_i(\mf x)$ splits $\L_{ii}$ into two different kinds of terms: The local terms at each of the interaction times and the ones that depend on the field correlations $\tau_i=\text{t}_1$ and at $\tau_i=\text{t}_2$. Namely,
    \begin{align}\label{eq:Lii}
        \L_{ii} =& P_i(\mf x_1)+P_i(\mf x_2)\nonumber\\
        &+ \cos(\Omega \Delta \text{t})\Re\ev{\hat{\phi}(\mf x_1)\hat{\phi}(\mf x_2)}\\
        &+ \sin(\Omega \Delta\text{t})\Im\ev{\hat{\phi}(\mf x_1)\hat{\phi}(\mf x_2)}\nonumber
    \end{align}
    where $\Delta\text{t} = \text{t}_2 - \text{t}_1$ and $P_i(\mf x)$ denotes the excitation probability of the detector after a delta coupled interaction at the event $\mf x$.
    
    Now, if one {wishes} to obtain the two-point function, one must subtract the probabilities for the single interactions $P_i(\mf x)$ from the total probability after the two interactions. To do this we require that the two-point function can be considered to be the same during the total time duration of the experiment\footnote{This assumption is not very restrictive in many relevant setups, such as in quantization schemes associated with time translation symmetries when there is a timelike Killing vector field. The reason is that if one performs an experiment and waits enough time for local perturbations to propagate away one can repeat measurements in (locally) identical conditions.}, where all the measurements necessary to obtain the the individual and joint probabilities occur, as argued in \cite{pipo}. Under these assumptions, it is possible to use Eq. \eqref{eq:Lii} to express the field's correlations in terms of experimentally accessible quantities:
    \begin{align}\label{eq:timelikeEq}
        \cos(\Omega \Delta \text{t})\Re&\ev{\hat{\phi}(\mf x_1)\hat{\phi}(\mf x_2)}+ \sin(\Omega \Delta\text{t})\Im\ev{\hat{\phi}(\mf x_1)\hat{\phi}(\mf x_2)}\nonumber\\
       & \!\!\!=\mathcal{L}_{ii}- P_i(\mf x_1) - P_i(\mf x_2).
    \end{align}
    Notice that in order to extract both the imaginary and real parts of the Wightman function between the events, it would be necessary to perform experiments with detectors of different energy gaps, so that $\Omega\Delta\text{t}$ can take the values of $n\pi$ and $(n+\frac{1}{2})\pi$, which will give the real and imaginary parts of the Wightman function, respectively~\cite{pipo}. Finally, notice that this derivation assumes pointlike localized detectors, but the same result can be obtained with sharply smeared detectors. As long as the smearing scale is much shorter than any of the other relevant scales, Eq.~\eqref{eq:timelikeEq} will still hold approximately. \textcolor{black}{This is not just `in principle' but rather a practical statement. Indeed, in Section \ref{sec:examples} we will show an example of how smeared detectors are able to accurately recover the spacetime metric from their correlations.}
    
    

    
    \subsubsection{Spacelike separated events} \label{sub:spacelike}
    
    In this section we will discuss how to extend the techniques of~\cite{pipo} to measure the two-point function of a quantum field for spacelike separated events by using pairs of detectors. In particular we will obtain the two-point function of the field in terms of the correlation function of probes that interact at different spacelike separated events. 
    
    We start by computing the correlation function of different probes whose interaction region is located around different {points}. First, notice that for $i\neq k$, we have
    \begin{align}
        \tr(\hat{\rho}_D \hat{\sigma}^+_i \hat{\sigma}^+_k) &= -\lambda^2 \mathcal{M}_{ik}\\
        \tr(\hat{\rho}_D \hat{\sigma}^-_i \hat{\sigma}^-_k) &= -\lambda^2 \mathcal{M}_{ik}^* \\
        \tr(\hat{\rho}_D \hat{\sigma}^+_i \hat{\sigma}^-_k) &= \lambda^2 \L_{ik}\\
        \tr(\hat{\rho}_D \hat{\sigma}^-_i \hat{\sigma}^+_k) &= \lambda^2 \L_{ik} = \lambda^2 \L_{ik}^*,
    \end{align}
    where $\mathcal{M}_{ik}\coloneqq \mathcal{N}_{ik} + \mathcal{N}_{ki}$. With this we can compute the correlation function for the monopole moments $\hat{\mu}_i$ at a given spatial slice $\Sigma_0$ defined by \mbox{$t = t_0$}. At time $t_0$, the $i$th detector will have a value of its proper time \mbox{$\tau_i=\tau_i^0$}, which corresponds to the point of their trajectory that lies in $\Sigma_0$. 
    Using $\mu_i(\tau_i^0) = e^{\ii\Omega\tau_i^0}\s_i^+ + e^{-\ii\Omega\tau_i^0}\s_i^-$, the correlation function between detectors $i$ and $k$ gives
    \begin{align}\label{eq:CLM}
        \mathcal{C}(i,k) &= \tr(\r_\textsc{d} \hat{\mu}_i(\tau_i^0)\hat{\mu}_k(\tau_k^0))\\&= 2\lambda^2\Re(e^{\ii\Omega(\tau_i^0-\tau_k^0)}\mathcal{L}_{ik} - e^{\ii\Omega (\tau_i^0+\tau_k^0)}\M_{ik}).\nonumber
    \end{align}
    This form of the detector-system two-point function is computed explicitly in Appendix \ref{app:corrfunc}.
    
\begin{figure}
        \centering
        \includegraphics[scale=0.35]{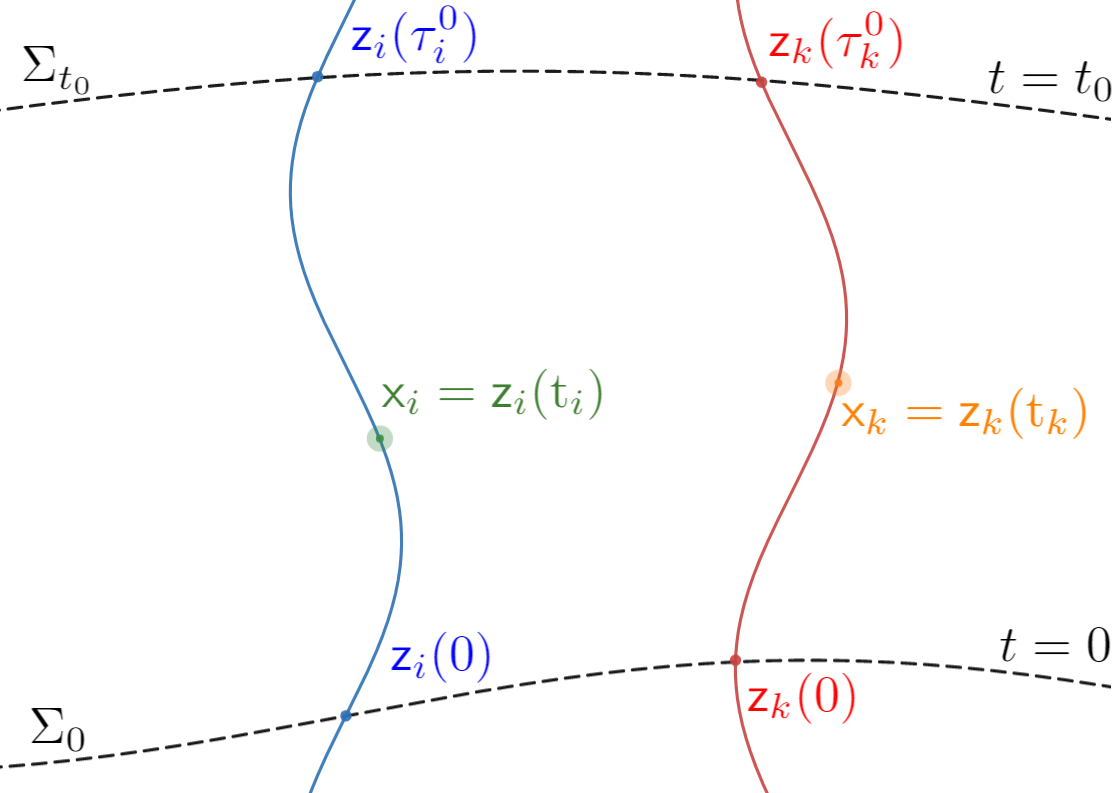}
        \caption{The setup described in Subsection \ref{sub:spacelike}, where two pointlike detectors interact between the slices $\Sigma_0$ and $\Sigma_{t_0}$.}
        \label{fig:diagram}
    \end{figure}

    
    We now reduce this expression to the case where each detector interacts with the quantum field with a delta coupling. The $i$th detector switches on its interaction at {the} point $\mf x_i$, such that $\mf x_i$ and $\mf x_k$ are spacelike separated for all $i$ {$\neq$} $k$. We label the proper time where the interaction takes place for each detector as $\text{t}_i=\tau_i(\mf x_i)$, such that $\mf z(\text{t}_i) = \mf x_i$ (See diagram in Fig.~\ref{fig:diagram}), so that it is possible to write the correlation function between the detectors in terms of the two-point function of the field. Formally, this corresponds to the choice of spacetime smearing \mbox{$\Lambda_i(\mf x) = \delta^{(n)}(\mf x - \mf x_i)/\sqrt{-g}$}. We obtain the following expression for the correlation function
    \begin{align}\label{eq:CikTwoPt}
        \mathcal{C}(i,k) \!&= 4\lambda^2\!\sin(\Omega(\text{t}_i \!+\! \tau_0))\sin(\Omega(\text{t}_k \!+\! \tau_0))\!\Re\!\ev{\!\hat{\phi}(\mf x_i)\hat{\phi}(\mf x_k)\!}\nonumber\\
        &= 4\lambda^2 \sin(\Omega(\text{t}_i \!+\!\tau_0))\sin(\Omega(\text{t}_k \!+\! \tau_0))\ev{\hat{\phi}(\mf x_i)\hat{\phi}(\mf x_k)},
    \end{align}
    where $\tau_0 = (\tau_i^0 +\tau_k^0)/2$. Notice that we can remove the `Re' from the equation since the two-point function is real. This is because the microcausality of the quantum field theory implies that the commutator of $\hat{\phi}(\mf{x})$ with itself at different spacelike separated points must vanish. In particular, this means that the detectors two-point correlator in Eq.~\eqref{eq:CikTwoPt} is proportional the field two-point function. This allows one to recover the two-point function of the {field} in terms of observables of the probe system for spacelike separated events.
    
    In the particular case of Minkowski spacetime, the setup described here takes a rather simple form. If the detectors are inertial and comoving, and the foliation $\Sigma_t$ is associated with their rest space, then $t$ can be chosen as their proper time parameter. If we further assume that both detectors interact at time $t = 0$ and are measured at a later time $t_0$, Eq. \eqref{eq:CikTwoPt} reads
    \begin{align}\label{eq:mee}
        \mathcal{C}(i,k) 
        &= 4\lambda^2 \sin^2(\Omega t_0)\ev{\hat{\phi}(\mf x_i)\hat{\phi}(\mf x_k)},
    \end{align}
    where here $\mf x_i$ and $\mf x_k$ are assumed to lie in the slice $\Sigma_0$. 
    
    In particular, for smeared detectors, one would expect that for sufficiently localized probes, \eqref{eq:mee} holds approximately. In this case, Eq. \eqref{eq:mee}, combined with Eq. \eqref{eq:CLM} implies that for any choice of measurement time $t_0$ in this setup, we have
    \begin{equation}\label{eq:Ct0}
        \sin^2(\Omega t_0)\ev{\hat{\phi}(\mf x_i)\hat{\phi}(\mf x_k)} \approx \frac{1}{2}\Re\left(\mathcal{L}_{ik} - e^{2 \ii\Omega t_0}\mathcal{M}_{ik}\right)
    \end{equation}
    in the limit where the size of the interaction region goes to {zero} and the spacetime smearing functions are thought to be nascent Dirac deltas. We can pick $t_0 = n\pi/\Omega$ for integer values of $n$, so that within the regime of validity of Eq.~\eqref{eq:Ct0}, we have
    \begin{equation}\label{eq:MisL}
        \Re\mathcal{L}_{ik} \approx \Re\mathcal{M}_{ik}.
    \end{equation}
    In order to explicitly write the field correlator in terms of the correlation function between the detectors, we use Eq. \eqref{eq:Ct0} again, but we now pick the measurement time $t_0$ as $t_0 = (n+\frac{1}{2})\pi/\Omega$, where $n$ is an integer. We then obtain a simple proportion between the detectors' correlation function at time $t_0$ and the correlation function of the the quantum field,
    \begin{align}\label{eq:spacelikeApprox}
       \ev{\hat{\phi}(\mf x_i)\hat{\phi}(\mf x_k)} &\approx \frac{1}{4\lambda^2}\mathcal{C}(i,k) \approx
        \Re\left(\mathcal{L}_{ik}\right),
    \end{align}
    where we used Eq. \eqref{eq:Ct0} and the approximation from Eq. \eqref{eq:MisL}. 
    
    Therefore, as a summary of this section, we have shown how it is possible to approximately recover the correlation function of the quantum field using sufficiently localized UDW detectors via Eqs.~\eqref{eq:timelikeEq} and~\eqref{eq:spacelikeApprox}.

    \section{Recovering the spacetime metric through local measurements of quantum fields}\label{sec:examples}

    In this section we will study in detail a setup that allows one to explicitly recover the spacetime metric by combining the results of Section \ref{sec:geometry} and \ref{sec:detectors}.

    \subsection{The general setup}\label{sub:general}
    
    We discussed in Section~\ref{sec:geometry} how one can obtain the spacetime metric from the Wightman function. On the other hand, the results of Section~\ref{sec:detectors} show that it is possible to obtain the exact form of the Wightman function in different points of {spacetime} if we have precise enough measurements of the correlations between UDW detectors. Putting these together, we can now show how it is possible to recover the spacetime metric from local measurements of quantum fields with particle detectors.
    
    
    We propose the following setup in order to recover the spacetime metric by locally measuring a quantum field in different {events}:
        \begin{enumerate}
            \item Couple local probes to a quantum field.
            \item Measure the correlation between the probes at different {spacetime points}.
            \item From the correlation between the detectors, compute the field two-point function between the corresponding events.
            \item Compute the metric by taking the coincidence limit in Eq. \eqref{mine}.
        \end{enumerate} 
    Although these steps might in principle seem simple, one must be careful with their implementation. Indeed, in order to obtain the spacetime metric with some precision, the limit of step four needs to be taken with enough precision. This relies on coupling probes separated by small enough spacetime intervals, so that the limit can be approximated well enough. In practice this requires significant control of the probe systems. However, in principle, it is possible to recover the spacetime metric with arbitrary precision using this procedure, provided that the probes are small enough and their coupling with the field can be regulated with enough precision.
    
    In the remaining subsections we will consider different spacetimes with UDW {detectors}. We will consider detectors that are either pointlike, or sufficiently small, that couple fast enough to the quantum field. This allows us to use the approximations \eqref{eq:timelikeEq} and \eqref{eq:spacelikeApprox}. We will compute the (experimentally accessible) correlation function between detectors placed in a local region of spacetime separated by a coordinate separation $L$ and adapt Eq.~\eqref{mine} to this discrete setup. In essence, given that the interaction happens very approximately pointlikewise in spacetime, we will effectively have access to the Wightman function associated to a discrete lattice of points. We then take the discrete derivative of the Wightman function using the points in the lattice given by the center of the interaction of the detectors\footnote{We will explicitly analyze the effect of a finite region of interaction in Subsection \ref{sub:smeared}.}.

    In our setup, we will assume that the experimentalist can use a coordinate system $x^\mu = (x^0,x^i)$ to label the events in spacetime where the measurements take place, but does not have access to any local notion of space or time separation. In other words, with this information it is possible to label events, but it is not possible to compute any physical spacetime distance.  Mathematically, this is  saying that spacetime can locally be regarded as a $D$-dimensional manifold, but that there is no known spacetime metric. Our goal is to use particle detectors that interact at events which are labelled with values of $x^\mu = (x^0,x^i)$, and from the read outs of those detectors, infer a spacetime metric components in the lab coordinates.

    We consider a set of $N^n$ detectors parametrized by $(\text{j}_1,...,\text{j}_n)$ where each $\text{j}_i$ runs from $1$ to $N$. For simplicity, let us work under the assumption that the detectors undergo trajectories associated to the coordinate system $x^\mu$, so that {they} move along the curves \mbox{$x^i = x^i_{\text{j}_i} = \text{const}$}. Then, $x^i_{\text{j}_i}$ are the constants that determine the spatial coordinates of each detector. This simplifying assumption will allow us to easily compute the metric in the $x^\mu$ coordinates. 
        
    \begin{figure}
        \centering
        \includegraphics[scale=0.43]{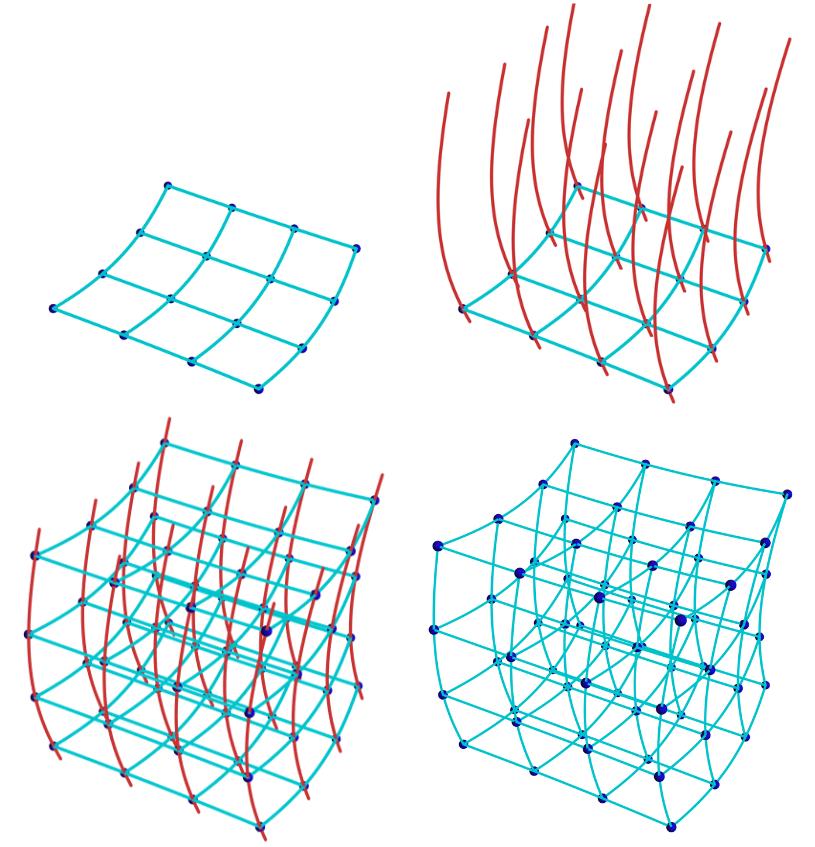}
        \caption{The setup described in Subsection \ref{sub:general}, where a lattice of particle detectors evolves through time and interacts with the field, effectively creating a spacetime lattice from the centres of interaction regions. }
        \label{fig:scheme}
    \end{figure}
    
    We consider that each detector interacts $N_0$ times with the field. The time coordinate of the center points of the interactions will be given by $x^0 = x^0_{\text{j}_0}$, with ${\text{j}_0}$ running from $1$ to $N_0$, corresponding to the values of time where the interactions happen. In this setup, we obtain a $D$-dimensional lattice of points labelled by $({\text{j}_0},\text{j}_1,...,\text{j}_n)$ associated to the events in which the detector interactions take place. A schematic representation of the setup can be found in Fig. \ref{fig:scheme}. We can then apply the formalism described in Section \ref{sec:explicit} by choosing the surfaces $\Sigma_t$ as the spacelike surfaces determined by $x^0 = \text{const}$. This allows one to obtain (from the readouts of the detectors) an approximation to the correlation function of the quantum field $W(\mf x,\mf x')$ when $\mf x$ and $\mf x'$ are events in the lattice. 

    As discussed, to recover the spacetime metric we will employ the discrete derivative of the Wightman function. We can obtain it directly from experimentally measurable detector data from the local measurements centered at the points parametrized by $\mathfrak{j}\coloneqq({\text{j}_0},\text{j}_1,...,\text{j}_n)$. We denote the coordinates of the interaction point $(x^0_{\text{j}_0},x^1_{\text{j}_1},...,x^n_{\text{j}_n})$ by $x^\mu_{\mathfrak{j}}$, so that after measuring the quantum field with the particle detectors, we obtain $W(\mf x_{\mathfrak{j}},\mf x_{\mathfrak{l}})$ for all values of the multi-indices $\mathfrak{j}$ and $\mathfrak{l}$.

    In order to write the discrete derivative in a simple way, we define the object
    \begin{equation}
        \bm 1_\mu = (\underbrace{0,...,0}_{{\mu-1}},1,0,...,0).
    \end{equation}
    With this convention and the labelling $x_{\mathfrak{j}}^\mu$ for the coordinates of the events, given a scalar function $f(\mf x)$, it is possible to write its discrete derivative as
    \begin{equation}\label{eq:wtf}
        \left. \pdv{f}{x^\mu} \right|_{\mf x =\mf x_{\mathfrak{j}}}  \approx \frac{f(\mf x_{\mathfrak{j}+\bm 1_\mu})-f(\mf x_{\mathfrak{j}})}{x^\mu_{\mathfrak{j}+\bm 1_\mu}-x^\mu_{\mathfrak{j}}}.
    \end{equation}
    Intuitively, Eq. \eqref{eq:wtf} compares the value of the function $f(\mf x)$ at nearby points and divides it by the step. In order to recover the spacetime metric, we will compute the derivatives of the function $(W(\mf x,\mf x'))^{\frac{2}{2-D}}$. Its discrete derivative at $(\mf x_{\mathfrak{j}},\mf x_{\mathfrak{l}})$ with respect to its different arguments can be written as
    \begin{widetext}
        \begin{equation}\label{eq:Wapprox}
            \left.\pdv{}{x^{\mu'}}\pdv{}{x^{\nu}}W^{\frac{2}{2-D}}(\mf x,\mf x')\right|_{\overset{\displaystyle{\mf x = \mf x_{\mathfrak{j}}}}{\displaystyle{\,\mf x'\!\!=\mf x_{\mathfrak{l}}}}}\approx\frac{ W^{\frac{2}{2-D}}(\mf x_{\mathfrak{j}+\bm 1_\nu},\mf x_{\mathfrak{l}+\bm 1_\mu})-W^{\frac{2}{2-D}}(\mf x_{\mathfrak{j}},\mf x_{\mathfrak{l}+\bm 1_\mu})-W^{\frac{2}{2-D}}(\mf x_{\mathfrak{j}+\bm 1_\nu},\mf x_{\mathfrak{l}})+W^{\frac{2}{2-D}}(\mf x_{\mathfrak{j}},\mf x_{\mathfrak{l}})}{(x^\mu_{\mathfrak{j}+\bm 1_\mu}-x^\mu_{\mathfrak{j}})(x^\nu_{\mathfrak{l}+\bm 1_\nu}-x^\nu_{\mathfrak{l}})}.
        \end{equation}    
    \end{widetext}
    This expression should give an approximate form for Eq. \eqref{mine}, so that we expect to recover the spacetime metric in the case where the detectors are separated by small enough values of {the coordinate separation} $x^\mu_{\mathfrak{j}+\bm 1_\mu}-x^\mu_{\mathfrak{j}}$.
    
    \textcolor{black}{To simplify the formalism for a proof of principle, we will assume the detectors to be separated by a coordinate distance $L$ in all directions (including the time direction). We can then rewrite Eq. \eqref{eq:Wapprox} using that the coordinates of $\mf x_{\mathfrak{j}+\bm 1_\mu}$ are $x_{\mathfrak{j}}^\mu+L\,\bm 1_\mu$. }
 It is important to remark that in this case, the parameter $L$ does not represent physical spacetime interval separation: it is merely a coordinate parameter. However, continuity ensures that when the coordinate separation between events go to zero, so does the spacetime interval between them. For this reason, Eq. \eqref{eq:Wapprox} will be used in the examples we study below, so that we assume that the detector coordinate separation is $L$ in all directions in the coordinate system that determines their trajectories. It is then expected that if $L$ is small enough, Eq. \eqref{eq:Wapprox} will yield a good approximation for the spacetime metric, once the numerical factor from Eq. \eqref{mine} is included. In fact, as we will see in the following examples, for pointlike detectors the metric will be precisely recovered when $L\rightarrow 0$, and very approximately recovered for smeared detectors when the distance between detectors approach the detectors size.
    
    \subsection{Example one: \tb{inertial} pointlike detectors in Minkowski spacetime}\label{sub:mink}
    
        In this subsection we consider the spacetime (unknown to the experimenter) to be (3+1) dimensional Minkowski, with a quantum field quantized according to an inertial frame
        . Then the quantum field can be written in terms of the creation and annihilation operators as
        \begin{equation}\label{eq:phimink}
            \hat{\phi}(\mf x) = \int \frac{\dd^3 \bm k}{\sqrt{2\omega_{\bm k}}}\left(\frac{e^{\ii \mf k \cdot \mf x}}{(2\pi)^{\frac{3}{2}}}\hat{a}_{\bm k}+\frac{e^{- \ii \mf k \cdot \mf x}}{(2\pi)^{\frac{3}{2}}}\hat{a}^\dagger_{\bm k}\right) .
        \end{equation}
    \begin{figure}[h!]
        \includegraphics[scale=0.7]{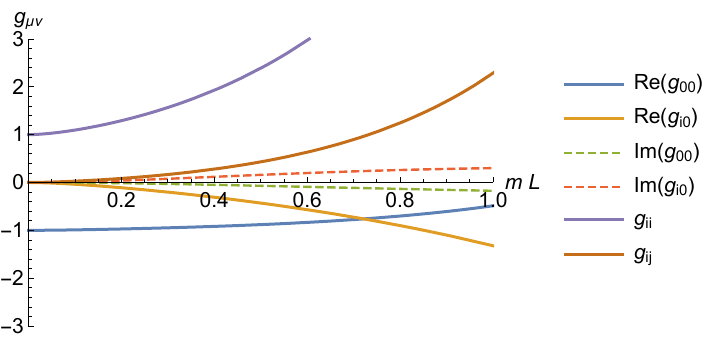}
        \caption{Estimation of the metric coefficients in terms of the coordinate distance between detectors, $L$, for inertial comoving detectors in Minkowski spacetime.}\label{fig:mink}
    \end{figure}
    
    We then consider the inertial coordinate system \mbox{$\mf x = (t,\bm x) \tb{= (t,x,y,z)}$}, and build the lattice of particle detectors in a local region of spacetime according to Section \ref{sec:detectors}. For simplicity, in this first example, we consider detectors that interact via delta couplings. In this case, we know it is possible to recover the Wightman function of the quantum field exactly, as was shown in Section \ref{sec:detectors}. We can then use Eq. \eqref{eq:Wapprox} to approximate the spacetime metric. We obtain the estimates for the metric shown in Figure \ref{fig:mink}. It is possible to see that the readouts of the detector approximate the metric coefficients as the distance between the detectors decreases. Moreover, the imaginary part of the approximate (experimentally obtained) metric goes to zero faster than the real components as $L\rightarrow 0$, so that we are only left with real expressions, which yield the expected value $g_{\mu\nu} = \text{diag}(-1,1,1,1)$.
    
    \color{black}
    \subsection{Example two: uniformly accelerated pointlike detectors in Minkowski spacetime}\label{sub:minkacc}
    
    In this section we consider uniformly accelerated pointlike detectors probing the Minkowski vacuum. The goal of this example is to see whether it is still possible to recover the spacetime metric in different coordinate systems built from particle detectors in different states of motion.  We then consider Rindler coordinates $(T,X,y,z)$ in Minkowski spacetime, associated to the inertial coordinates $(t,\bm x)$ from Subsection \ref{sub:mink} by
    \begin{equation}
        \begin{cases}
            t = X \sinh(a T),\\
            x = X \cosh(a T),
        \end{cases}
    \end{equation}
    with $X>0$ and $T\in\mathbb{R}$. The Minkowski line element in this coordinate system then reads
    \begin{equation}
        \dd s^2 = - a^2 X^2 \dd T^2 + \dd X^2 + \dd y^2 + \dd z^2.
    \end{equation}

    \begin{figure}[h!]
        \includegraphics[scale=0.7]{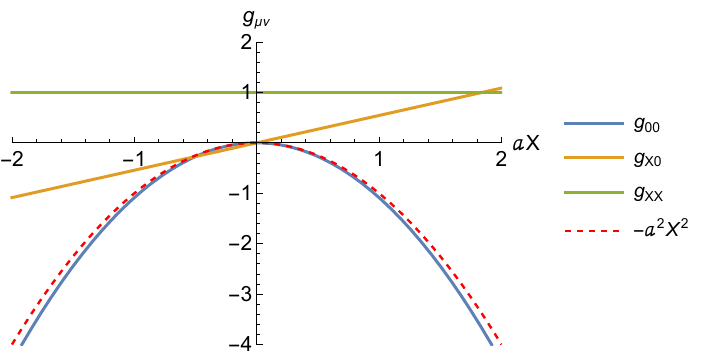}
        \includegraphics[scale=0.7]{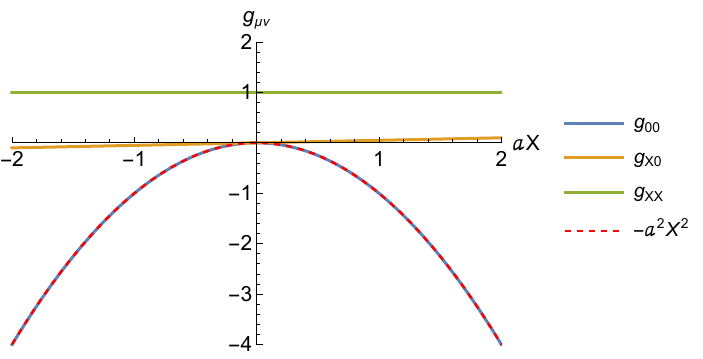}
        \caption{\tb{Metric coefficients obtained from the correlation function of the quantum field in Minkowski spacetime with accelerated detectors. The metric coefficients are plotted as a function of the coordinate $a X$ of the detectors. The detectors were separated by a coordinate distance {$L = a^{-1}$} in the top plot and $L = 0.1a^{-1}$ in the bottom plot.}}\label{fig:MinkAcc}
    \end{figure}

    The lattice of detectors which is associated to this coordinate system is such that each detector follows a trajectory defined by $X = \text{const.}$ with constant values of $y$ and $z$. That is, each detector is uniformly accelerated with different proper acceleration given by $1/X$. We consider a massless field, and detectors interacting along different Dirac deltas situated along the corresponding motions of the Rindler flow. Performing the computation in Eq. \eqref{eq:Wapprox}, we find the estimates for the spacetime metric shown in Fig. \ref{fig:MinkAcc} for metric as a function of the coordinate $X$ for detector separations of $L = a^{-1}$ in the top plot and $L = 0.1 a^{-1}$ in the bottom plot. 

    In the limit of $L = 0$ we recover the metric exactly, as would be expected. Overall, we recover the expected behaviour of the metric components with the coordinate distance $X$ between the detectors. The smaller the value of $L$, the better the fit between the curves. Also notice that for higher values of $aX$, we find more discrepancies between the estimated metric components and the actual Minkowski metric. This is due to the fact that the time separation between the interactions is proportional to $aX$. Overall, we find that it is possible to recover the spacetime metric even when the detectors are in different states of motion, giving rise to different coordinate systems which express the same spacetime metric. This is a general feature of the setup we have considered: it is generally covariant, so that regardless of the relative motion of the detectors, one can recover the metric in the coordinate system associated with their trajectories.

    \color{black}
    
    \subsection{Example three: hyperbolic static Robertson-Walker spacetime}
    
    Consider the hyperbolic cosmological spacetime with a constant scale factor $a$. Then the metric in comoving coordinates coordinates can be written as
    \begin{equation}
        \dd s^2 = -\dd t^2 + a^2(\dd \chi^2 + \sinh^2(\chi)(\dd \theta^2 + \sin^2\theta\dd \phi^2)),
    \end{equation}
    We then reparametrize it using the conformal time parameter $\eta = t/a$, so that the coordinates read
    \begin{equation}
        \dd s^2 = a^2(-\dd \eta^2 + \dd \chi^2 + \sinh^2(\chi)(\dd \theta^2 + \sin^2\theta\dd \phi^2)).
    \end{equation}
    Quantizing a conformally coupled real scalar quantum field $\hat{\phi}(\mf x)$ with respect to the conformal time, we can expand it in terms of creation and annihilation operators,
    \begin{align}
        \hat{\phi}(\mf x) = \sum_{k=1}^\infty \sum_{l=0}^{k-1}& \sum_{m=-l}^l \frac{1}{\sqrt{2a^2\omega_{\bm k}}}\\
        &\left(e^{-\ii\omega_{\bm k}\eta}\Pi_{kl}^{(-)}(\chi)Y_l^m(\theta,\phi)\hat{a}_{\bm k}+\text{H.c.}\right)\nonumber
    \end{align}
    \begin{figure}[h!]
        \includegraphics[scale=0.7]{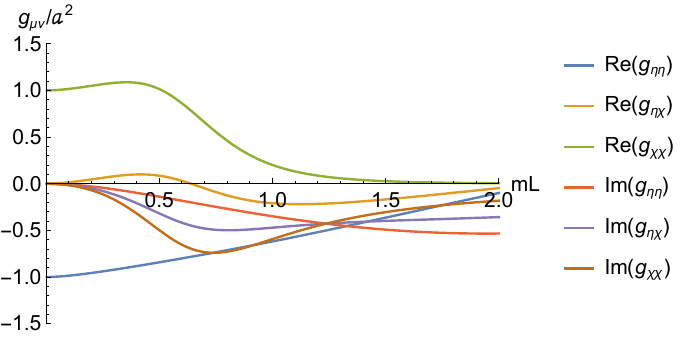}
        \caption{Metric coefficients $g_{\eta\eta}$, $g_{\eta \chi}$, $g_{\chi\eta}$ and $g_{\chi\chi}$ in terms of the coordinate distance between detectors, $L$ in the hyperbolic static Robertson Walker spacetime with choices of mass and conformal parameters such that $\mu = a m$.}\label{fig:cosmo}
    \end{figure}
    where $\omega_{\bm k} = k^2 + \mu^2$ and $\mu^2 = a^2(m^2 + (\xi -\frac{1}{6})R)$, where $R$ is the Ricci scalar, which is constant in this spacetime. The explicit expression for $\Pi^{(-)}_{kl}(\chi)$ can be found in e.g. \cite{birrell_davies}. The vacuum Wightman function can then be explicitly computed and reads
    \begin{equation}
        W(\mf x,\mf x') = \frac{\ii\mu(\chi-\chi')H_1^{(2)}(\mu[(\eta-\eta')^2 - (\chi-\chi')^2])}{8\pi a^2\sinh(\chi-\chi')[(\eta-\eta')^2 - (\chi-\chi')^2]},
    \end{equation}
    where $H_1^{(2)}$ is the Hankel function.

    The results of particle detectors separated by a coordinate distance $L$ in the $(\eta, \chi)$ coordinates coupled to this spacetime can be found in Figure \ref{fig:cosmo}. We then see that in the limit of $L\rightarrow 0$, we recover the exact metric coefficients.

    \subsection{Example four: deSitter spacetime}
    
    In this example we recover the metric of four-dimensional deSitter spacetime by probing it with particle detectors. deSitter spacetime has a constant curvature with scalar curvature, $R = \text{const.}>0$. It is then possible to write the Riemann curvature tensor as
    \begin{equation}
        R_{\mu \nu \rho \sigma} = \frac{1}{\ell^2}\qty(g_{\mu \rho} g_{\nu \sigma} - g_{\mu \sigma} g_{\nu \rho}),
    \end{equation}
    where $\ell$ is the curvature radius of the spacetime. This will be the first example we investigate where the metric components explicitly depend on the coordinates we use to prescribe the detector's trajectories. 
    \begin{figure}[h]
        \includegraphics[scale=0.7]{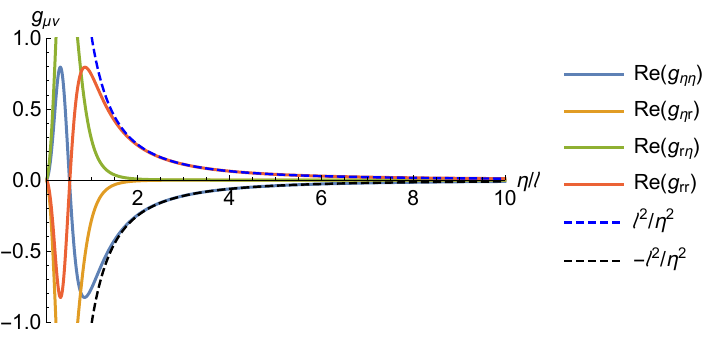}
        \includegraphics[scale=0.7]{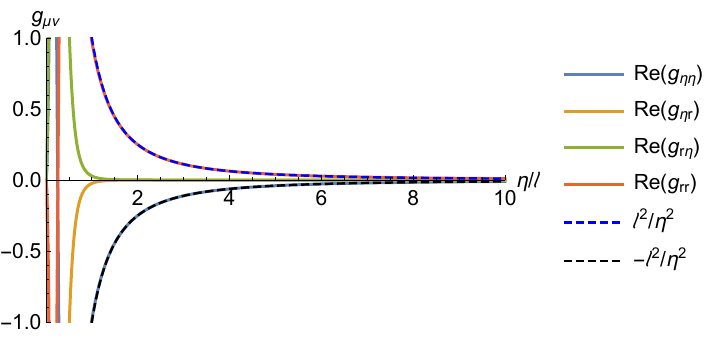}
        \caption{Metric coefficients calculated from the correlation function of the quantum field in deSitter spacetime. The metric coefficients are plotted as a function of the coordinate $\eta/\ell$ of the detectors and we choose $\nu = 9/4$. The detectors were separated by a coordinate distance {$L = e^{-7.5}\ell$} in the top plot and $L = e^{-7.5} \ell/2$ in the bottom plot.}\label{fig:deSitter}
    \end{figure}
    
    We consider conformal coordinates in deSitter spacetime, so that the metric can be written as
    \begin{equation}
        \dd s^2 = \frac{\ell^2}{\eta^2}\left(-\dd \eta^2 + \dd x^2 +\dd y^2 + \dd z^2\right).
    \end{equation}
    The quantization of a real scalar field with respect to the modes adapted to this coordinate system yields the following vacuum Wightman function~\cite{birrell_davies}:
    \begin{align}
        W(\mf x,\mf x') = &\frac{1}{16\pi \ell^2}\left(\frac{1}{4}-\nu^2\right)\sec(\pi \nu)\\
        &\times{}_2F_1\left(\frac{3}{2}+\nu,\frac{3}{2}-\nu,2,1+\frac{(\Delta \eta)^2-|\Delta \bm x|^2}{4 \eta'\eta}\right),\nonumber
    \end{align}
    where ${}_2F_1$ is the Hypergeometric function and we write $\mf x = (\eta,\bm x)$ and $\mf x'= (\eta',\bm x')$. We define $\Delta\eta = \eta - \eta'$ and $\Delta \bm x = \bm x - \bm x'$. The parameter $\nu$ contains the information regarding the mass of the field and its coupling to curvature. It is explicitly given by
    \begin{equation}
        \nu^2 = \frac{9}{4}-12\left(\frac{m^2}{R}+\xi \right).
    \end{equation}
    
    In order to recover the metric in this spacetime, we consider delta-coupled particle detectors that undergo trajectories defined by $\bm x = \text{const}.$, separated by a coordinate distance $L$. We consider these detectors to interact at conformal times which are multiples of $L$, as detailed in Subsection \ref{sub:general}. In Fig. \ref{fig:deSitter}, we plot the metric approximation for two values of $L$ as a function of $\eta$. As expected, when $L\rightarrow 0$, we approximate the function $\pm \ell^2/\eta^2$ with high precision. Also notice that the method yields better approximations for larger values of $\eta/\ell$. This is due to the fact that at a given fixed value of conformal time $\eta$, the proper space separation between neighbouring detector trajectories is given by $\frac{\ell}{\eta} L$, which is smaller for larger values of $\eta/\ell$.
    
    \color{black}
    
    \subsection{Example five: The half Minkowski space with Dirichlet boundary conditions}
    
    In this example we study the effect of boundary conditions in our protocol for recovering the spacetime metric. We analyze a massless Klein-Gordon field in the half Minkowski space $\mf x = (t,x,y,z)$ with $z\geq 0$ and Dirichlet boundary conditions at $z = 0$. This effectively restricts the basis of solutions for the Klein Gordon equation and changes the field's two point function. The vacuum state that respects the symmetries of this spacetime then yields the Wightman function~\cite{birrell_davies,alex}
    \begin{equation}
        W(\mf x,\mf x') = \frac{1}{8\pi^2}\frac{1}{\sigma} - \frac{1}{8\pi^2}\frac{1}{\sigma_*},
    \end{equation}
    where $\sigma = \sigma(\mf x,\mf x')$ and $\sigma_* = \sigma_*(\mf x,\mf x')$ are given by
    \begin{align}
        \sigma &= \frac{1}{2}\left(-(t-t')^2 + (x-x')^2 + (y-y')^2 + (z-z')^2\right),\nonumber\\
        \sigma_* &= \frac{1}{2}\left(-(t-t')^2 + (x-x')^2 + (y-y')^2 + (z+z')^2\right).
    \end{align}
    Then, it is possible to verify that whenever $\mf x$ or $\mf x'$ lies at the plane $z = 0$, $W(\mf x,\mf x') = 0$, as expected.
    
    \begin{figure}[h!]
        \includegraphics[scale=0.7]{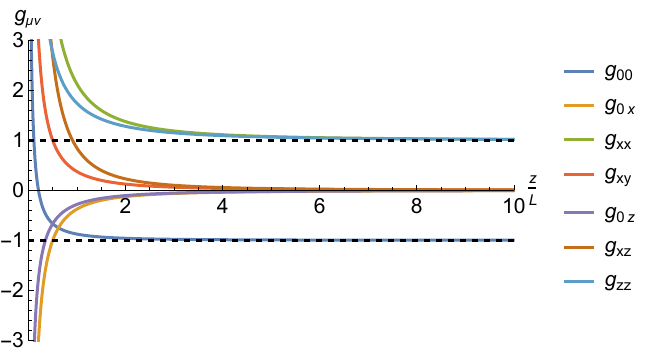}
        \caption{\tb{Metric coefficients calculated from the correlation function of a massless quantum field in the half Minkowski space with Dirichlet boundary conditions at $z=0$. The metric coefficients are plotted as a function of the coordinate ratio between their $z$ coordinate and the separation between the detectors, $L$.}}\label{fig:Minkz}
    \end{figure}
    
    We then consider pointlike particle detectors at rest with respect to the spatial part of the $\mf x = (t,x,y,z)$ coordinate system separated by a coordinate distance $L$ which interact with the quantum field at events separated in coordinate time by $L$. Then, following the procedure outlined in Subsection \ref{sub:general}, we estimate the metric coefficients using Eq. \eqref{eq:Wapprox}. In Fig. \ref{fig:Minkz} we plot the obtained metric coefficients as a function of the ration between the coordinate distance $z$ and the separation between the detectors. As we see, the further away from the boundary, the better the metric estimation is. Moreover, due to the fact that $W(\mf x,\mf x') = 0$ at the boundary, the computation of Eq. \eqref{eq:Wapprox} yields a divergent result at $z = 0$, showing that at the boundary, it is not possible to estimate the metric coefficients. Nevertheless, we highlight that for any $z>0$, the limit $L \longrightarrow 0$ yields the exact Minkowski metric coefficients. 
    
    Overall, we see that the presence of a boundary disturbs the metric estimation, and fails at the boundary itself. Nevertheless, for any point that is not at the boundary, the correlation function of particle detectors can be used to accurately yield the metric of spacetime same as in the previous cases.
    
    \color{black}
    
    \subsection{Example six: one-particle Fock states in Minkowski spacetime}
    
    In this example we consider one-particle Fock wavepackets in Minkowski spacetime instead of the vacuum to show with an example how the recovery of the spacetime geometry is independent of the field state. We consider the same setup as that of Subsection \ref{sub:mink}, with the expansion of Eq. \eqref{eq:phimink}. With respect to these modes, a general normalized one-particle state $\ket{\psi}$ can be written as
    \begin{equation}
        \ket{\psi} = \int \dd^3 \bm k \,f(\bm k) \hat{a}^\dagger_{\bm k} \ket{0},
    \end{equation}
    where $f$ is an $L^2(\mathbb{R}^3)$ normalized function. The two-point function of the field in the state $\ket{\psi}$ will be given by (see, e.g.,~\cite{antiparticles})
    \begin{equation}
        W_\psi(\mf x,\mf x') =  W_0(\mf x,\mf x') + F(\mf x) F^*(\mf x') + F(\mf x') F^*(\mf x),
    \end{equation}
    where $W_0(\mf x,\mf x')$ is the Minkowski vacuum Wightman function and
    \begin{equation}
        F(\mf x) = \frac{1}{(2\pi)^{\frac{3}{2}}}\int \frac{\dd^3 \bm k}{\sqrt{2 \omega_{\bm k}}} \, f(\bm k) e^{\ii \mf k \cdot\mf x}.
    \end{equation}

    \begin{figure}[h]
        \includegraphics[scale=0.7]{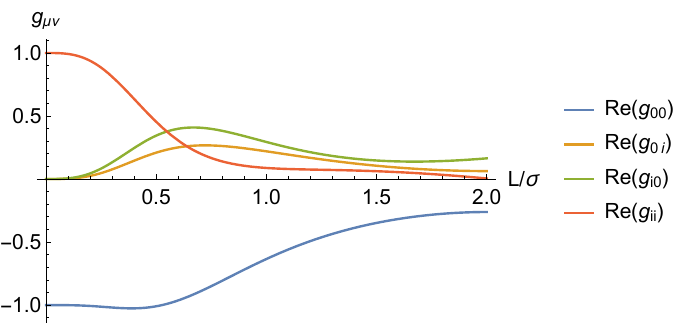}
        \caption{{Estimation of metric} coefficients {obtained using} particle detectors in Minkowski spacetime when a massless field is in a Gaussian one-particle state. The metric coefficients are plotted in terms of the coordinate distance between detectors, $L$.}\label{fig:part}
    \end{figure}
    
    For a concrete example, we consider the field to be massless ($m=0$) and prescribe the momentum profile function that defines $\psi$ as a Gaussian centred at $\bm k = \bm 0$ with standard deviation $\sigma$,
    \begin{equation}
        f(\bm k) = \frac{1}{(\pi\sigma)^{3/4}} e^{-\frac{\bm k^2}{2 \sigma^2}}.
    \end{equation}
    We use delta-coupled detectors interacting in events separated by a time/space coordinate separation of $L$. Figure \ref{fig:part} shows the value of the approximated metric coefficients obtained from the detector measurements as a function of the separation between detector interaction events. This allows us to recover the Minkowski metric in the limit where $L \rightarrow 0$. We also find that for larger values of $L$ the results {begin to show} state dependence (compare Figs. \ref{fig:mink} and \ref{fig:part}). This is expected since it is only when the detectors are close to each other that the measurements converge to the coefficients of the metric independently of the state.

    \color{black}
    \subsection{Example seven: smeared detectors probing the Minkowski vacuum}\label{sub:smeared}

    In this subsection we consider the example of non-pointlike inertial detectors probing the vacuum of Minkowski spacetime in order to recover the spacetime metric. Unlike the point-like case, it is not possible to recover the Wightman function of the quantum field exactly using smeared particle detectors. However, if the detectors are small, we can resort to the approximation pointed out in Eq. \eqref{eq:spacelikeApprox}. Although it is expected that smeared detectors will provide a less accurate measurement of the spacetime metric, these models represent realistic physical systems that are not infinitely localized, such as, for example, atoms interacting with the electromagnetic field~\cite{eduardo,Jonsson1,Nicho1,nadine,nadine,richard}.
     \begin{figure}[h!]
        \includegraphics[scale=0.8]{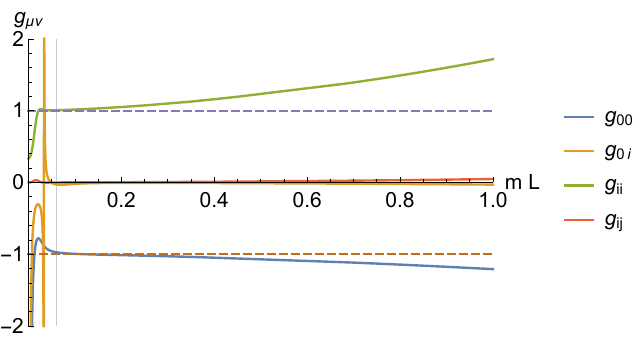}
        \caption{Metric coefficients extracted by gaussian smeared particle detectors in Minkowski spacetime when the field is in the vacuum state. We have chosen $\Omega = m$, where $m$ is the mass of the field and $\sigma = 10^{-2} \Omega^{-1}$. The metric coefficients are plotted in terms of the proper distance between detectors, $L$. The vertical line on the left indicates $L = 6\sigma$}\label{fig:ohmy}
    \end{figure}
    \begin{figure}[h!]
        \includegraphics[scale=0.76]{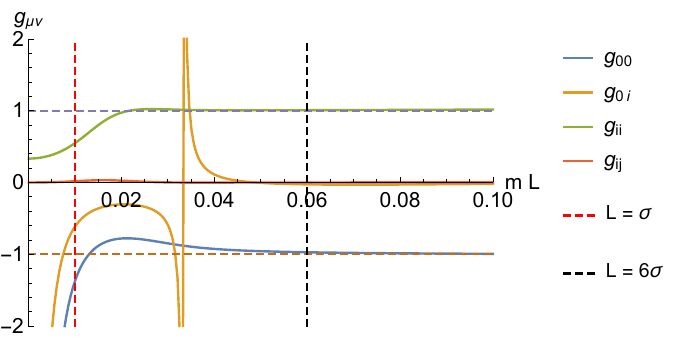}
        \caption{Metric coefficients extracted by gaussian smeared particle detectors in Minkowski spacetime when the field is in the vacuum state. We have chosen $\Omega = m$, where $m$ is the mass of the field and $\sigma = 10^{-2} \Omega^{-1}$. The metric coefficients are plotted in terms of the proper distance between detectors, $L$.}\label{fig:ohmyzoom}
    \end{figure}
    
    In this example we consider a lattice of inertial Gaussian-smeared detectors labelled by $\mathfrak{j}$, whose interactions are centered at sites $\mf x_{\mathfrak{j}}$ such that spacetime smearing function can be written as
    \begin{equation}
        \Lambda_{\text{j}}(\mf x) = \sum_{\text{j}_0=1}^{N_0}\frac{e^{-\frac{(\mf x - \mf x_{\mathfrak{j}})^2}{2\sigma^2}}}{(2\pi\sigma^2)^{D/2}},
    \end{equation}
    where $(\mf x - \mf x_{\mathfrak{j}})^2 = \delta_{\mu\nu} (x^\mu - x_{\mathfrak{j}}^\mu)(x^\nu - x_{\mathfrak{j}}^\nu)$. For each detector trajectory, we sum over the different interaction times $\text{j}_0$. Notice that this corresponds to an interaction that lasts for a time equal to the light crossing time of the detector's spatial profile. We consider such interaction because Eq. \eqref{eq:spacelikeApprox} is only expected to be valid in the limit $\sigma\rightarrow 0$, where we obtain \mbox{$\Lambda(\mf x)\longrightarrow\delta(\mf x-\mf x_{\mathfrak{j}})$}. This limit is only true if the spatio-temporal profile of the interaction are approximately equal.
    
    In Fig. \ref{fig:ohmy} we plot the approximated metric obtained from detector measurements as a function of their coordinate separation $L$, when one considers the approximate Wightman function from Eq.~\eqref{eq:spacelikeApprox}.  For detector separations smaller or comparable to $5\sigma$, the detectors' spacetime smearings have a significant overlap, which makes this regime unphysical. In Fig.~\ref{fig:ohmyzoom} we show a scaled version of the plot, where the approximated metric is shown for smaller values of $L$, and the spurious behaviour for small $L$. Nevertheless, the metric is accurately recovered when $L$ is between $5\sigma$ and $10\sigma$ even when one considers smeared detectors.
    
    \color{black}

    \section{Conclusions}\label{sec:conclusions}
    
    We showed how one can recover the spacetime metric through local measurements of quantum fields. More concretely, we generalized the results of \cite{pipo}, showing that it is possible to use particle detectors to obtain the two-point function of a quantum field evaluated at both timelike and spacelike separated events. Armed with this knowledge, we mapped the (measurable) correlations acquired by arrays of particle detectors to the two-point function of the field. With this data---and inspired by the results of Kempf et al.~\cite{achim,achim2}---we were able to accurately recover the spacetime metric as the detectors separations become sufficiently small. We also showed how due to the fact that all (reasonable) states behave similarly to the vacuum at short scales, our result is valid for any physical state of the quantum field.
    
    We have fully analyzed several explicit examples where pointlike detectors were used to recover the metric in flat and curved spacetimes. Namely, we studied Minkowski spacetime, the hyperbolic static Robertson Walker spacetime and deSitter spacetime. In all cases we found that as the detector separation goes to zero, we recover the exact metric at the location of interaction from local measurements on the detector observables. We also studied the case of  finite-sized detectors in Minkowski spacetime, where we found that it is possible to recover the metric in the limit where the detector separation is small enough, but larger than the detectors size. Even with Gaussian smearings, when the separation is larger than $5\sigma$, we are able to recover the metric coefficients with great precision.
    
    In summary, we were able to obtain a notion of space and time intervals based on the measurement outcomes of (in principle experimentally realizable) quantum particle detectors. This shares the philosophy of recent work on the formulation of quantum reference frames~\cite{qrfReview,f19,f20a,f20,f21}, where the notions of space and time separations between observers are formulated in terms of shared quantum resources.
    
    Finally, as outlook, we conjecture that these explorations together with the pioneering work in~\cite{achim,achim2}  could pave the way for a formulation of causal structure  exclusively in terms of local quantum probes. These quantum particle detectors are the measurement devices that should perhaps replace the rulers and clocks of Einstein's relativity in scales where classical rulers or clocks stop making sense, and provide some  insight on the microscopic structure of what we call spacetime. 
    
    
    \section{Acknowledgements}\label{sec:Ack}

    T.R.P. and E.M.M. thank Achim Kempf and Flaminia Giacomini for insightful discussions. Research at Perimeter Institute is supported in part by the Government of Canada through the Department of Innovation, Science and Industry Canada and by the Province of Ontario through the Ministry of Colleges and Universities. E. M-M. is funded by the NSERC Discovery program as well as his Ontario Early Researcher Award.

\appendix

\onecolumngrid

\section{The pointlike limit of the correlation functions}\label{app:corrfunc}

In this Appendix we show Eqs. \eqref{eq:Lii} and Eq. \eqref{eq:CLM}  from Section \ref{sec:explicit} that express the field correlation in terms of measurable detector observables.

We start by considering the case where the field correlator is evaluated at timelike separated points analyzed {in} Subsection \ref{sub:timelike}. In this setup we consider a single probe interacting with the quantum field twice at spacetime points $\mf x_1$ and $\mf x_2$ such that $\tau_i(\mf x_1) = \text{t}_1$ and $\tau_i(\mf x_2) = \text{t}_2$. The corresponding spacetime smearing function is given by $\Lambda_i(\mf x) = \delta^{(4)}(\mf x - \mf x_1)/ \sqrt{-g}+\delta^{(4)}(\mf x - \mf x_2)/ \sqrt{-g}$. We can then compute the term $\mathcal{L}_{ii}$ from Eq. \eqref{eq:LiiFirst}:
\begin{align}
    \mathcal{L}_{ii} &= \int \!\!\dd V \dd V'\Lambda_i(\mf x)\Lambda_i(\mf x') e^{-\ii\Omega(\tau_i-\tau_i')}\!\!\ev{\hat{\phi}(\mf x)\hat{\phi}(\mf x')}\\&= \int \!\!\dd V \dd V' (\delta^{(4)}(\mf x - \mf x_1)/ \sqrt{-g}+\delta^{(4)}(\mf x - \mf x_2)/ \sqrt{-g})(\delta^{(4)}(\mf x' - \mf x_1)/ \sqrt{-g}+\delta^{(4)}(\mf x' - \mf x_2)/ \sqrt{-g}) e^{-\ii\Omega(\tau_i-\tau_i')}\!\!\ev{\hat{\phi}(\mf x)\hat{\phi}(\mf x')}\nonumber \\
    &= \ev{\hat{\phi}(\mf x_1)\hat{\phi}(\mf x_1)}+\ev{\hat{\phi}(\mf x_2)\hat{\phi}(\mf x_2)}+ e^{-\ii\Omega \Delta \text{t}}\!\!\ev{\hat{\phi}(\mf x_2)\hat{\phi}(\mf x_1)}+ e^{\ii\Omega \Delta \text{t}}\!\!\ev{\hat{\phi}(\mf x_1)\hat{\phi}(\mf x_2)}\nonumber\\
    &=P_i(\mf x_1)+P_i(\mf x_2)+ \cos(\Omega \Delta \text{t})\Re\ev{\hat{\phi}(\mf x_2)\hat{\phi}(\mf x_1)}+ \sin(\Omega \Delta \text{t})\Im\ev{\hat{\phi}(\mf x_1)\hat{\phi}(\mf x_2)},\nonumber
\end{align}
where $\Delta \text{t} = \text{t}_2 - \text{t}_1$ and $P_i(\mf x)$ denotes the delta coupling interaction of detector $i$ with the field at spacetime point $\mf x$. Also notice that $\ev{\hat{\phi}(\mf x_2)\hat{\phi}(\mf x_1)} = \ev{\hat{\phi}(\mf x_1)\hat{\phi}(\mf x_2)}^*$, so that $\Re\ev{\hat{\phi}(\mf x_2)\hat{\phi}(\mf x_1)} = \Re\ev{\hat{\phi}(\mf x_1)\hat{\phi}(\mf x_2)}$.

We have studied the relationship between the field two-point function and the detector correlators for spacelike events in Subsection \ref{sub:spacelike}. In this setup we consider two detectors that interact at spacelike separated regions, according to the setup depicted in Fig. \ref{fig:diagram} and detailed in Subsection \ref{sub:spacelike}. Plugging in the explicit expressions for $\mathcal{L}_{ik}$ and $\mathcal{N}_{ik}$ from Eq. \eqref{eq:Lik} we obtain:
\begin{align}
    \Re\left(e^{\ii\Omega(\tau_i^0-\tau_k^0)}\mathcal{L}_{ik} - e^{\ii\Omega (\tau_i^0+\tau_k^0)}(\mathcal{N}_{ik} + \mathcal{N}_{ki})\right)=\Re\left(e^{\ii\Omega(\tau_i^0-\tau_k^0)}\mathcal{L}_{ik} - e^{\ii\Omega (\tau_i^0+\tau_k^0)}\mathcal{N}_{ik} -e^{-\ii\Omega (\tau_i^0+\tau_k^0)} \mathcal{N}_{ki}^*\right).
\end{align}
We then work with the second expression. We have
\begin{align}
    &e^{\ii\Omega(\tau_i^0-\tau_k^0)}\mathcal{L}_{ik} - e^{\ii\Omega (\tau_i^0+\tau_k^0)}(\mathcal{N}_{ik} + \mathcal{N}_{ki})  \\&= \int \dd V \dd V' \Lambda_i(\mf x)\Lambda_k(\mf x') \ev{\hat{\phi}(\mf x)\hat{\phi}(\mf x')}\Big(e^{-\ii\Omega(\tau_i(\mf x) - \tau_k(\mf x'))}-e^{\ii\Omega(\tau_i^0+\tau_k^0) }e^{\ii\Omega(\tau_i(\mf x) + \tau_k(\mf x'))}\theta(t-t')\nonumber\\
    &\:\:\:\:\:\:\:\:\:\:\:\:\:\:\:\:\:\:\:\:\:\:\:\:\:\:\:\:\:\:\:\:\:\:\:\:\:\:\:\:\:\:\:\:\:\:\:\:\:\:\:\:\:\:\:\:\:\:\:\:\:\:\:\:\:\:\:\:\:\:\:\:\:\:\:\:\:\:\:\:\:\:\:\:\:\:\:\:\:\:\:\:\:\:\:\:\:\:\:\:\:\:\:\:\:\:\:\:\:\:\:\:\:\:\:\:\:\:\:\:\:\:\:-e^{-\ii\Omega (\tau_i^0+\tau_k^0)}e^{-\ii\Omega(\tau_i(\mf x) + \tau_k(\mf x'))}\theta(t'-t)\Big)\nonumber\\
    &= 2\ii\int \dd V \dd V'\Lambda_i(\mf x)\Lambda_k(\mf x')\ev{\hat{\phi}(\mf x)\hat{\phi}(\mf x')}\Big(\sin(\Omega (\tau_k(\mf x')+\tau_0))e^{-\ii\Omega (\tau_i(\mf x)+\tau_0)}\theta(t' - t)\nonumber\\
    &\:\:\:\:\:\:\:\:\:\:\:\:\:\:\:\:\:\:\:\:\:\:\:\:\:\:\:\:\:\:\:\:\:\:\:\:\:\:\:\:\:\:\:\:\:\:\:\:\:\:\:\:\:\:\:\:\:\:\:\:\:\:\:\:\:\:\:\:\:\:\:\:\:\:\:\:\:\:\:\:\:\:\:\:\:\:\:\:\:\:\:\:\:\:\:\:\:\:\:\:\:\:\:\:\:\:\:\:\:\:\:\:\:\:\:\:\:\:\:\:\:\:\: -\sin(\Omega (\tau_i(\mf x)+\tau_0))e^{\ii\Omega (\tau_k(\mf x')+\tau_0)}\theta(t-t')\Big)\nonumber,
\end{align}
where we used $1 = \theta(t-t') + \theta(t' - t)$ and we have defined
\begin{equation}
    \tau_0 \coloneqq \frac{\tau_i^0 + \tau_k^0}{2}.
\end{equation}
When each detector interacts with a symmetric delta coupling, there is no need to time order the exponential that defines the time evolution operator and we can simply replace $\theta(t-t')\rightarrow 1/2$ (See Appendix D of \cite{PozasDegenerate}). This gives us an expression for the equal time correlation function above when we choose $\Lambda(\mf x) = \delta^{(4)}(\mf x - \mf x_i)/ \sqrt{-g}$, denoting $\tau_i(\mf x_i) = \text{t}_i$. Namely, we obtain:
\begin{align}
    e^{\ii\Omega(\tau_i^0-\tau_k^0)}\mathcal{L}_{ik} - e^{\ii\Omega (\tau_i^0+\tau_k^0)}(\mathcal{N}_{ik} + \mathcal{N}_{ki}) &= \ii\left(\sin(\Omega (\text{t}_k+\tau_0))e^{-\ii\Omega (\text{t}_i+\tau_0)}-\sin(\Omega (\text{t}_i+\tau_0))e^{\ii\Omega (\text{t}_k+\tau_0)}\right)\ev{\hat{\phi}(\mf x_i)\hat{\phi}(\mf x_k)}\nonumber\\
    &= 2\sin(\Omega(\text{t}_i + \tau_0))\sin(\Omega(\text{t}_k + \tau_0))\ev{\hat{\phi}(\mf x_i)\hat{\phi}(\mf x_k)},
\end{align}
which establishes the result of Eq. \eqref{eq:CikTwoPt}.

\section{State independence of the short distance limit behaviour of the Wightman function}\label{app:stateind}

In this appendix we show that the behaviour of the Wightman function of a free scalar field in the coincidence limit is independent of the state of the field. As an intermediate step, we will also show that the two-point correlator of any regular enough field state can be written as the vacuum Wightman function plus state dependent terms.

Specifically, we will show for any normalized pure state $\ket{\psi}$, we have that
\begin{equation}
    W_\psi(\mf x,\mf x') = \bra{\psi} \hat{\phi}(\mf x) \hat{\phi}(\mf x') \ket{\psi} = W(\mf x,\mf x')+ \sum_{m=0}^\infty F_m(\mf x) G^*_m(\mf x')+\text{H.c.},
\end{equation}
where $F_m(\mf x)$ and $G_m(\mf x')$ are state dependent regular functions in the limit $\mf x'\rightarrow \mf x$ and $W(\mf x,\mf x') = \bra{0}\! \hat{\phi}(\mf x) \hat{\phi}(\mf x')\!\ket{0}$ is the vacuum Wightman function, hence the behaviour of $ W_\psi(\mf x,\mf x')$ in this limit is independent of the state of the field. Although the coincidence limit is formally divergent, we show is that the singular part of $W_{\psi}(\mf x,\mf x')$ is the same as that of the vacuum Wightman function $W(\mf x,\mf x')$. Notice that by showing this result for pure state, the result also follows for arbitrary normalized mixed states, as these are given by convex combinations of pure states. 

A general pure state $\ket{\psi}$ in the Fock space associated to a quantum field $\hat{\phi}(\mf x)$, in the quantization scheme discussed in Subsection \ref{sub:KG} is given by
\begin{equation}
    \ket{\psi} = \sum_{m=0}^\infty \int \dd^n {\bm k}_1 ... \dd^n {\bm k}_m  f_{m}(\bm {\bm k}_1,..., \bm {\bm k}_m) \hat{a}_{\bm {\bm k}_1}^\dagger... \hat{a}_{\bm {\bm k}_m}^\dagger \ket{0},
\end{equation}
where the functions $f_m$ define the momentum profile of the $m$ particle content of the state. Notice that due to the bosonic nature of the field, the functions $f_m$ are symmetric with respect to all of their arguments. Also notice that the function $f_0$ is a constant, which allows for superpositions between the particle sectors and the vacuum state. 

It is possible to find a condition for the $L^2$ norm of the functions $f_m$ using the following result,
\begin{equation}\label{eq:comm}
    \bra{0} \hat{a}_{\bm p_r}... \hat{a}_{\bm p_1}\hat{a}_{\bm {\bm k}_1}^\dagger... \hat{a}_{\bm {\bm k}_m}^\dagger\ket{0} = \delta_{rm}\sum_{\sigma\in S_m} \delta(\bm {\bm k}_1 - \bm p_{\sigma(1)})... \delta(\bm {\bm k}_m - \bm p_{\sigma(m)}),
\end{equation}
where $S_m$ denotes the set of all permutations of $m$ elements. With this, we have that
\begin{equation}\label{eq:normalization}
    \braket{\psi}{\psi} = 1 \Rightarrow  \sum_{m=0}^\infty  m!\int \dd^n {\bm k}_1 ... \dd^n {\bm k}_m |f_{m}(\bm k_1,..., \bm k_m)|^2 = 1.
\end{equation}
This condition will be important later to show that the two-point function of any state can be written as the vacuum Wightman function added to a regular term.

We proceed to compute the two-point function on $\ket{\psi}$ explicitly. Using the expansion of the free scalar quantum field from Eq. \eqref{eq:modeExp} we have
\begin{align}
    &\bra{\psi} \hat{\phi}(\mf x) \hat{\phi}(\mf x') \ket{\psi} = \int \dd^3 \bm p_0 \dd^3 \bm {\bm k}_0 \bra{\psi} \left(u_{\bm p_0}(\mf x)\hat{a}_{\bm p_0} + u^*_{\bm p_0}(\mf x)\hat{a}^\dagger_{\bm p_0} \right)\left(u_{\bm k_0}(\mf x')\hat{a}_{\bm k_0} + u^*_{\bm k_0}(\mf x')\hat{a}^\dagger_{\bm k_0} \right)\ket{\psi}\\
    &= \sum_{m,r = 0}^\infty  \int \dd^3 {\bm p}_0 \dd^3 {\bm p}_1...\dd^3 {\bm p}_r\dd^3 {\bm k}_0\dd^3 {\bm k}_1...\dd^3 {\bm k}_m f_{r}^*(\bm p_1,..., \bm p_r)f_{m}(\bm {\bm k}_1,..., \bm {\bm k}_m)\nonumber\\
    &\:\:\:\:\:\:\:\:\:\:\:\:\times\bra{0} \hat{a}_{\bm p_1}...\hat{a}_{\bm p_r} \left(u_{\bm p_0}(\mf x)\hat{a}_{\bm p_0} + u^*_{\bm p_0}(\mf x)\hat{a}^\dagger_{\bm p_0} \right)\left(u_{\bm {\bm k}_0}(\mf x')\hat{a}_{\bm {\bm k}_0} + u^*_{\bm {\bm k}_0}(\mf x')\hat{a}_{\bm {\bm k}_0}^\dagger\right) \hat{a}_{\bm {\bm k}_1}^\dagger... \hat{a}_{\bm {\bm k}_m}^\dagger \ket{0}\nonumber\\
    &= \sum_{m,r = 0}^\infty  \int \dd^3 {\bm p}_0 \dd^3 {\bm p}_1...\dd^3 {\bm p}_r\dd^3 {\bm k}_0 \dd^3 {\bm k}_1...\dd^3 {\bm k}_m f_{r}^*(\bm p_1,..., \bm p_r)f_{m}(\bm k_1,..., \bm k_m)\label{eq:openTwoPt}\\
    &\:\:\:\:\:\:\:\:\:\:\:\:\times \Big(u_{\bm p_0}(\mf x)u_{\bm k_0}(\mf x')\bra{0} \hat{a}_{\bm p_1}...\hat{a}_{\bm p_r} \hat{a}_{\bm p_0}\hat{a}_{\bm k_0} \hat{a}_{\bm k_1}^\dagger... \hat{a}_{\bm k_m}^\dagger \ket{0}+u^*_{\bm p_0}(\mf x)u_{\bm k_0}(\mf x')\bra{0} \hat{a}_{\bm p_1}...\hat{a}_{\bm p_r}\hat{a}^\dagger_{\bm p_0} \hat{a}_{\bm k_0} \hat{a}_{\bm k_1}^\dagger... \hat{a}_{\bm k_m}^\dagger \ket{0}\nonumber\\
    &\:\:\:\:\:\:\:\:\:\:\:\:\:\:\:\:\:\:\:\:\:\:\:\:+u_{\bm p_0}(\mf x)u^*_{\bm k_0}(\mf x')\bra{0} \hat{a}_{\bm p_1}...\hat{a}_{\bm p_r}\hat{a}_{\bm p_0} \hat{a}_{\bm k_0}^\dagger \hat{a}_{\bm k_1}^\dagger... \hat{a}_{\bm k_m}^\dagger \ket{0}
    +u^*_{\bm p_0}(\mf x)u^*_{\bm k_0}(\mf x')\bra{0} \hat{a}_{\bm p_1}...\hat{a}_{\bm p_r}\hat{a}^\dagger_{\bm p_0}\hat{a}_{\bm k_0}^\dagger \hat{a}_{\bm k_1}^\dagger... \hat{a}_{\bm k_m}^\dagger \ket{0}\Big).\nonumber
\end{align}
Let us compute each of the matrix elements that show up above separately. The third term is immediately computed from Eq. \eqref{eq:comm}. It yields
\begin{equation}
    \bra{0} \hat{a}_{\bm p_1}...\hat{a}_{\bm p_r}\hat{a}_{\bm p_0} \hat{a}_{\bm k_0}^\dagger \hat{a}_{\bm k_1}^\dagger... \hat{a}_{\bm k_m}^\dagger \ket{0} = \delta_{rm} \sum_{\sigma\in S_{m+1}}\delta(\bm k_1 - \bm p_{\sigma(1)})...\delta(\bm k_{m+1} - \bm p_{\sigma(m+1)}),
\end{equation}
where we identify $ 0 =m+1$ for the permutations above. The first and last matrix elements in Eq. \eqref{eq:openTwoPt} can also be computed by means of Eq. \eqref{eq:comm}. That is, if we denote $\bm p_0 = \bm p_{r+1}$ and $\bm k_0 = \bm p_{r+2}$, we have
\begin{equation}
    \bra{0} \hat{a}_{\bm p_1}...\hat{a}_{\bm p_r}\hat{a}_{\bm p_0} \hat{a}_{\bm k_0} \hat{a}_{\bm k_1}^\dagger... \hat{a}_{\bm k_m}^\dagger \ket{0} = \delta_{m, r+2} \sum_{\sigma\in S_{r+2}}\delta(\bm k_1 - \bm p_{\sigma(1)})...\delta(\bm k_{m} - \bm p_{\sigma(m)}).
\end{equation}
Regarding the fourth term, if we now denote $\bm k_0 = \bm p_{m+1}$ and $\bm p_0 = \bm k_{m+2}$, we have,
\begin{equation}\label{eq:comm2}
    \bra{0} \hat{a}_{\bm p_1}...\hat{a}_{\bm p_r}\hat{a}_{\bm p_0}^\dagger \hat{a}_{\bm k_0}^\dagger \hat{a}_{\bm k_1}^\dagger... \hat{a}_{\bm k_m}^\dagger \ket{0} = \delta_{r, m+2} \sum_{\sigma\in S_{m+2}}\delta(\bm k_0 - \bm p_{\sigma(0)})...\delta(\bm k_{m} - \bm p_{\sigma(m)}).
\end{equation}
Finally, for the second term in Eq. \eqref{eq:openTwoPt}, we must use the canonical commutation relations in order to obtain
\begin{align}
    \bra{0} \hat{a}_{\bm p_1}...\hat{a}_{\bm p_r}&\hat{a}^\dagger_{\bm p_0} \hat{a}_{\bm k_0} \hat{a}_{\bm k_1}^\dagger... \hat{a}_{\bm k_m}^\dagger \ket{0} = -\delta(\bm k_0 - \bm p_0)\bra{0} \hat{a}_{\bm p_1}...\hat{a}_{\bm p_r} \hat{a}_{\bm k_1}^\dagger... \hat{a}_{\bm k_m}^\dagger \ket{0} + \bra{0} \hat{a}_{\bm p_1}...\hat{a}_{\bm p_r} \hat{a}_{\bm k_0}\hat{a}^\dagger_{\bm p_0} \hat{a}_{\bm k_1}^\dagger... \hat{a}_{\bm k_m}^\dagger \ket{0}\\
    &=-\delta(\bm k_0 - \bm p_0) \delta_{rm}\sum_{\sigma\in S_m} \delta(\bm k_1 - \bm p_{\sigma(1)})...\delta(\bm k_m - \bm p_{\sigma(m)})+\delta_{rm}\sum_{\sigma\in S_{m+1}} \delta(\bm k_1 - \bm p_{\sigma(1)})...\delta(\bm k_{m+1} - \bm p_{\sigma(m+1)}),\nonumber
\end{align}
where in the second term we are denoting $\bm k_0 = \bm p_{r+1}$ and $\bm p_0 = \bm k_{m+1}$. Noticing that the first term above corresponds to the permutations on $S_{m+1}$ that leave $m+1$ fixed, we can rewrite the whole term as
\begin{align}
    \bra{0} \hat{a}_{\bm p_1}...\hat{a}_{\bm p_r}\hat{a}^\dagger_{\bm p_0} \hat{a}_{\bm k_0} \hat{a}_{\bm k_1}^\dagger... \hat{a}_{\bm k_m}^\dagger \ket{0} = \delta_{rm}\sum_{\sigma\in S^{\text{last}}_{m+1}} \delta(\bm k_1 - \bm p_{\sigma(1)})...\delta(\bm k_{m+1} - \bm p_{\sigma(m+1)}),
\end{align}
where $S_{m+1}^{\text{last}}$ denotes the set of permutations of $m+1$ elements that \emph{do not} leave the last term fixed.

Notice that the fact that $f_m$ are symmetric with respect to its arguments can be translated into the statement that
\begin{equation}
    f_m(\bm k_1,...,\bm k_m) = 
    f_m(\bm k_\sigma(1),...,\bm k_\sigma(m)),
\end{equation}
whenever $\sigma\in S_m$. We then have:
\begin{align}
    &\sum_{m,r = 0}^\infty  \int \dd^3 {\bm p}_0 \dd^3 {\bm p}_1...\dd^3 {\bm p}_r\dd^3 {\bm k}_0\dd^3 {\bm k}_1...\dd^3 \bm k_m f_{r}^*(\bm p_1,..., \bm p_r)f_{m}(\bm k_1,..., \bm k_m) u_{\bm p_0}(\mf x)u_{\bm {\bm k}_0}(\mf x')\bra{0} \!\hat{a}_{\bm p_1}...\hat{a}_{\bm p_r} \hat{a}_{\bm p_0}\hat{a}_{\bm k_0} \hat{a}_{\bm k_1}^\dagger... \hat{a}_{\bm k_m}^\dagger \!\ket{0}\nonumber\\
    &=\!\!\!\!\!\sum_{m,r = 0}^\infty \int \!\!\dd^3 {\bm p}_{r+1} \dd^3 {\bm p}_1...\dd^3 {\bm p}_r\dd^3 {\bm p}_{r+2}\dd^3 {\bm k}_1...\dd^3 {\bm k}_m f_{r}^*(\bm p_1,..., \bm p_r)f_{m}(\bm k_1,..., \bm k_m) u_{\bm p_0}(\mf x)u_{\bm k_0}(\mf x')\bra{0} \!\hat{a}_{\bm p_1}...\hat{a}_{\bm p_r} \hat{a}_{\bm p_0}\hat{a}_{\bm k_0} \hat{a}_{\bm k_1}^\dagger... \hat{a}_{\bm k_m}^\dagger\! \ket{0}\nonumber\\
    &=\sum_{r = 0}^\infty  \int \dd^3 {\bm p}_{r+1} \dd^3 {\bm p}_1...\dd^3 {\bm p}_r\dd^3 {\bm p}_{r+2}\dd^3 {\bm k}_1...\dd^3 {\bm k}_{r+2} f_{r}^*(\bm p_1,..., \bm p_r)f_{r+2}(\bm k_1,..., \bm k_{r+2})\\
    &\:\:\:\:\:\:\:\:\:\:\:\:\:\:\:\:\:\:\:\:\:\:\:\:\:\:\:\:\:\:\:\:\:\:\:\:\:\:\:\:\:\:\:\:\:\:\:\:\:\:\:\:\:\:\:\:\:\:\:\:\:\:\:\:\:\:\:\:\:\:\:\:\:\:\:\:\:\:\:\:\:\:\:\:\:\:\:\:\:\:\:\:\:\:\:\:\:\:\:\:\:\:\:\:\:\:\:\:\times u_{\bm p_{r+1}}(\mf x)u_{\bm p_{r+2}}(\mf x')\sum_{\sigma\in S_{r+2}}\delta(\bm k_1 - \bm p_{\sigma(1)})...\delta(\bm k_{r+2} - \bm p_{\sigma(r+2)})\nonumber\\
    &=\sum_{r = 0}^\infty\sum_{\sigma\in S_{r+2}}  \int \dd^3 {\bm p}_{r+1} \dd^3 {\bm p}_1...\dd^3 {\bm p}_r\dd^3 {\bm p}_{r+2}\dd^3 {\bm k}_1...\dd^3 {\bm k}_{r+2} f_{r}^*(\bm p_1,..., \bm p_r)f_{r+2}(\bm k_1,..., \bm k_{r+2})\nonumber\\
    &\:\:\:\:\:\:\:\:\:\:\:\:\:\:\:\:\:\:\:\:\:\:\:\:\:\:\:\:\:\:\:\:\:\:\:\:\:\:\:\:\:\:\:\:\:\:\:\:\:\:\:\:\:\:\:\:\:\:\:\:\:\:\:\:\:\:\:\:\:\:\:\:\:\:\:\:\:\:\:\:\:\:\:\:\:\:\:\:\:\:\:\:\:\:\:\:\:\:\:\:\:\:\:\:\:\:\:\:\times u_{\bm p_{r+1}}(\mf x)u_{\bm p_{r+2}}(\mf x')\delta(\bm k_{\sigma(1)} - \bm p_{1})...\delta(\bm k_{\sigma(r+2)} - \bm p_{r+2})\nonumber\\
    &=\sum_{r = 0}^\infty\sum_{\sigma\in S_{r+2}}  \int \dd^3 {\bm p}_{r+1} \dd^3 {\bm p}_1...\dd^3 {\bm p}_r\dd^3 {\bm p}_{r+2}\dd^3 \bm  k_{\sigma(1)}...\dd^3 \bm k_{\sigma(r+2)} f_{r}^*(\bm p_1,..., \bm p_r)f_{r+2}(\bm k_{\sigma(1)},..., \bm k_{\sigma(r+2)})\nonumber\\
    &\:\:\:\:\:\:\:\:\:\:\:\:\:\:\:\:\:\:\:\:\:\:\:\:\:\:\:\:\:\:\:\:\:\:\:\:\:\:\:\:\:\:\:\:\:\:\:\:\:\:\:\:\:\:\:\:\:\:\:\:\:\:\:\:\:\:\:\:\:\:\:\:\:\:\:\:\:\:\:\:\:\:\:\:\:\:\:\:\:\:\:\:\:\:\:\:\:\:\:\:\:\:\:\:\:\:\:\:\times u_{\bm p_{r+1}}(\mf x)u_{\bm p_{r+2}}(\mf x')\delta(\bm k_{1} - \bm p_{1})...\delta(\bm k_{r+2} - \bm p_{r+2})\nonumber\\
    &=\sum_{r = 0}^\infty\sum_{\sigma\in S_{r+2}}  \int \dd^3 {\bm p}_{r+1} \dd^3 {\bm p}_1...\dd^3 {\bm p}_r\dd^3 {\bm p}_{r+2}\dd^3 {\bm k}_{1}...\dd^3 {\bm k}_{r+2} f_{r}^*(\bm p_1,..., \bm p_r)f_{r+2}(\bm k_1,..., \bm k_{r+2})\nonumber\\
    &\:\:\:\:\:\:\:\:\:\:\:\:\:\:\:\:\:\:\:\:\:\:\:\:\:\:\:\:\:\:\:\:\:\:\:\:\:\:\:\:\:\:\:\:\:\:\:\:\:\:\:\:\:\:\:\:\:\:\:\:\:\:\:\:\:\:\:\:\:\:\:\:\:\:\:\:\:\:\:\:\:\:\:\:\:\:\:\:\:\:\:\:\:\:\:\:\:\:\:\:\:\:\:\:\:\:\:\:\times u_{\bm p_{r+1}}(\mf x)u_{\bm p_{r+2}}(\mf x')\delta(\bm k_{1} - \bm p_{1})...\delta(\bm k_{r+2} - \bm p_{r+2})\nonumber\\
    &=\sum_{r = 0}^\infty(r+2)!\sum_{\sigma\in S_{r+2}}  \int \dd^3 {\bm p}_1...\dd^3 {\bm p}_r\dd^3 {\bm p}_{r+2} \dd^3 {\bm p}_{r+1}u_{\bm p_{r+1}} (\mf x)u_{\bm p_{r+2}}(\mf x') f_{r}^*(\bm p_1,..., \bm p_r)f_{r+2}(\bm p_1,..., \bm p_{r+2}).\nonumber
\end{align}
In the chain of equalities above, we have, respectively: 1) relabelled ${\bm p}_0$ as ${\bm p}_{r+1}$ and ${\bm k}_0$ as ${\bm p}_{r+2}$. 2) Used the result from Eq. \eqref{eq:comm2}. 3) Dragged the sum over $\sigma$ to outside the integral and made the change of variables over $\sigma$, $\sigma \rightarrow \sigma^{-1}$ so that the permutations now act on the ${\bm k}_i$ argument of the Dirac delta functions. 4) Made the change of variables in the integrals over ${\bm k}_p$, ${\bm k}_p \rightarrow {\bm k}_{\sigma(p)}$. 5) Used symmetry of $f_m$ in its arguments to get rid of the permutations over $\sigma$. 6) Performed the integration over the Dirac delta functions. Notice how the assumed of $f_m$ allows us to ``contract'' the modes with any argument of $f_{r+2}$ and simplify the result. Notice that in the equation above, the mode functions always show up multiplied by integrable functions, which implies that the term above is regular.


The contribution to the Wightman function from the second term reads
\begin{align}\label{eq:secTerm}
    &\sum_{m,r = 0}^\infty  \int \dd^3 {\bm k}_0 \dd^3 {\bm k}_1...\dd^3 {\bm k}_r\dd^3 {\bm k}_0\dd^3 {\bm k}_1...\dd^3 {\bm k}_m f_{r}^*(\bm p_1,..., \bm p_r)f_{m}(\bm k_1,..., \bm k_m)\\
    &\:\:\:\:\:\:\:\:\:\:\:\:\:\:\:\:\:\:\:\:\:\:\:\:\:\:\:\:\:\:\:\:\:\:\:\:\:\:\:\:\:\:\:\:\:\:\:\:\:\:\:\:\:\:\:\:\:\:\:\:\:\:\:\:\:\:\:\:\:\:\:\:\:\:\:\:\:\:\:\:\:\times  u^*_{\bm p_0}(\mf x) u_{\bm k_0}(\mf x')\delta_{rm}\sum_{\sigma\in S^{\text{last}}_{m+1}} \delta(\bm k_1 - \bm p_{\sigma(1)})...\delta(\bm k_{m+1} - \bm p_{\sigma(m+1)})\nonumber\\
    &= 
    \sum_{m = 0}^\infty \sum_{\sigma\in S_{m+1}^{\text{last}}} \int \dd^3 {\bm k}_{m+1} \dd^3 {\bm k}_1...\dd^3 {\bm k}_m\dd^3 p_{m+1}\dd^3 {\bm k}_1...\dd^3 {\bm k}_m f_{m}^*(\bm p_1,..., \bm p_m)f_{m}(\bm k_1,..., \bm k_m)\nonumber\\
    &\:\:\:\:\:\:\:\:\:\:\:\:\:\:\:\:\:\:\:\:\:\:\:\:\:\:\:\:\:\:\:\:\:\:\:\:\:\:\:\:\:\:\:\:\:\:\:\:\:\:\:\:\:\:\:\:\:\:\:\:\:\:\:\:\:\:\:\:\:\:\:\:\:\:\:\:\:\:\:\:\:\times  u^*_{\bm k_{m+1}}(\mf x) u_{\bm p_{m+1}}(\mf x') \delta(\bm k_1 - \bm p_{\sigma(1)})...\delta(\bm k_{m+1} - \bm p_{\sigma(m+1)})\nonumber\\
    &= 
    \sum_{m = 0}^\infty \sum_{\sigma\in S_{m+1}^{\text{last}}} \int \dd^3 {\bm p}_1...\dd^3 {\bm p}_m\dd^3 {\bm p}_{m+1}  u_{\bm p_{m+1}}(\mf x')u^*_{\bm p_{\sigma(m+1)}}(\mf x) f_{m}^*(\bm p_1,..., \bm p_m)f_{m}(\bm p_{\sigma(1)},..., \bm p_{\sigma(m)}).\nonumber
\end{align}
Notice that in the expression above 
the permutations in $S_{m+1}^{\text{last}}$ are precisely those that \emph{do not} leave the last term fixed. This implies that the functions $u^*_{\bm p_{\sigma(m+1)}}(\mf x)$ and $u_{\bm p_{m+1}}(\mf x')$ will never be contracted with each other (the term ``contracted'' here refers to integrals over the repeated momentum variables in the product of two functions). 
Together with the normalization condition for the functions $f_m$, this implies that all terms in Eq. \eqref{eq:secTerm} are regular functions of $\mf x$ and $\mf x'$. 

The third term in Eq. \eqref{eq:openTwoPt} can be written as
\begin{align}
    &\sum_{m,r = 0}^\infty  \int \dd^3 {\bm k}_0 \dd^3 {\bm k}_1...\dd^3 {\bm k}_r\dd^3 {\bm k}_0\dd^3 {\bm k}_1...\dd^3 {\bm k}_m f_{r}^*(\bm p_1,..., \bm p_r)f_{m}(\bm k_1,..., \bm k_m)\\
    &\:\:\:\:\:\:\:\:\:\:\:\:\:\:\:\:\:\:\:\:\:\:\:\:\:\:\:\:\:\:\:\:\:\:\:\:\:\:\:\:\:\:\:\:\:\:\:\:\:\:\:\:\:\:\:\:\:\:\:\:\:\:\:\:\:\:\:\:\:\:\:\:\:\:\:\:\:\:\:\:\:\times u_{\bm p_0}(\mf x)u^*_{\bm k_0}(\mf x')\delta_{rm} \sum_{\sigma\in S_{m+1}}\delta(\bm k_1 - \bm p_{\sigma(1)})...\delta(\bm k_{m+1} - \bm p_{\sigma(m+1)})\nonumber\\
    &=\sum_{m= 0}^\infty \sum_{\sigma\in S_{m+1}} \int \dd^3 {\bm k}_{m+1} \dd^3 {\bm k}_1...\dd^3 {\bm k}_m f_{m}^*(\bm p_1,..., \bm p_m)f_{m}(\bm k_{\sigma(1)},..., \bm p_{\sigma(m)})u_{\bm p_{m+1}}(\mf x)u^*_{\bm k_{m+1}}(\mf x').\nonumber
\end{align}
Now notice that in the equation above we only have that the modes are contracted with each other in the terms that leave $m+1$ fixed. This allows us to split the sum over the permutations as
\begin{equation}
    \sum_{\sigma\in S_{m+1}} = \sum_{\sigma\in S_m} + 
    \sum_{\sigma\in S_{m+1}^{\text{last}}}.
\end{equation}
The terms that involve $S_{m+1}^{\text{last}}$ are related to the matrix element of Eq. \eqref{eq:secTerm} by conjugation, while the terms associated with the permutations in $S_m$ are given by
\begin{align}
    &\sum_{m = 0}^\infty  \sum_{\sigma\in S_m}\int \dd^3 {\bm k}_{m+1} \dd^3 {\bm k}_1...\dd^3 {\bm k}_mf_{m}^*(\bm p_1,..., \bm p_r)f_{m}(\bm p_{\sigma(1)},..., \bm p_{\sigma(m)})u^*_{\bm p_{m+1}}(\mf x) u_{\bm k_{m+1}}(\mf x')\delta(\bm k_{m+1} - \bm p_{m+1})\\
    &=\sum_{m = 0}^\infty  \sum_{\sigma\in S_m}\int \dd^3 {\bm k}_{m+1} \dd^3 {\bm k}_1...\dd^3 {\bm k}_mf_{m}^*(\bm p_1,..., \bm p_r)f_{m}(\bm p_{1},..., \bm p_{m})u^*_{\bm p_{m+1}}(\mf x) u_{\bm k_{m+1}}(\mf x')\delta(\bm k_{m+1} - \bm p_{m+1})\nonumber\\
    &= \int \dd^3 {\bm k}_0\dd^3 {\bm k}_0 u^*_{\bm {\bm k}_0}(\mf x) u_{\bm {\bm k}_0}(\mf x')\delta(\bm k_0 - \bm p_0) = \int \dd^3 {\bm k}_0 u^*_{\bm p_0}(\mf x) u_{\bm k_0}(\mf x') \nonumber\\
    &= W_0(\mf x,\mf x'),\nonumber
\end{align}
where we have used the symmetry of the $f_m$ terms and the normalization condition from Eq. \eqref{eq:normalization}.

We have thus shown that in a quantization scheme associated to modes as in Eq. \eqref{eq:modeExp}, the Wightman function in
\emph{any} pure state can be written as the vacuum Wightman function added to regular terms in the limit $\mf x'\rightarrow \mf x$, which also generalizes to mixed states. In particular, this means that the major contribution will be due to the divergences in the vacuum Wightman function, and the other terms become negligible in the $\mf x \rightarrow \mf x'$ limit.

\twocolumngrid

\bibliography{references}

\end{document}